\documentclass[11pt]{elsarticle}
\biboptions{sort&compress,numbers}
\usepackage{microtype}
\usepackage[margin=1in]{geometry}
\usepackage{amsmath,amssymb,amsthm,bm,stmaryrd}
\usepackage{subcaption}
\biboptions{sort&compress}
\usepackage{upgreek}
\usepackage{epstopdf}
\usepackage{float}
\usepackage{xcolor}
\usepackage[labelfont=bf]{caption}
\usepackage[export]{adjustbox}
\AppendGraphicsExtensions{.tif}

\usepackage[final]{changes} 
\definecolor{C0}{HTML}{1F77B4}
\definecolor{C1}{HTML}{FF7F0E}
\definecolor{C2}{HTML}{2ca02c}
\definecolor{C3}{HTML}{d62728}
\definecolor{C4}{HTML}{9467bd}
\definecolor{C5}{HTML}{8c564b}
\definechangesauthor[color=C0]{R1} 
\definechangesauthor[color=C1]{R2} 
\definechangesauthor[color=C3]{RA} 

\begin{document}

    \begin{frontmatter}

        \title{Investigating shock wave propagation, evolution, and anisotropy using a moving window concurrent atomistic-continuum framework}
        \author[auburn]{Alexander S. Davis}
        \author[auburn]{Vinamra Agrawal\corref{cor1}}
        \ead{vinagr@auburn.edu}
        \cortext[cor1]{Corresponding author}
        \address[auburn]{Department of Aerospace Engineering, Auburn University, Auburn, AL USA}
        
        \begin{abstract}
            Despite their success in microscale modeling of materials, atomistic methods are still limited by short time scales, small domain sizes, and high strain rates.
            Multiscale formulations can capture the continuum-level response of solids over longer runtimes, but using such schemes to model highly dynamic, nonlinear phenomena is very challenging and an active area of research.
            In this work, we develop novel techniques within the concurrent atomistic-continuum multiscale framework to simulate shock wave propagation through a two-dimensional, single-crystal lattice.
            The technique is described in detail, and two moving window methods are incorporated to track the shock front through the domain and thus prevent spurious wave reflections at the atomistic-continuum interfaces.
            We compare our simulation results to analytical models as well as previous atomistic and CAC data and discuss the apparent effects of lattice orientation on the shock response of FCC crystals.
            We then use the moving window techniques to perform parametric studies which analyze the shock front's structure and planarity. 
            Finally we compare the efficiency of our model to molecular dynamics simulations.
            This work showcases the power of using a moving window concurrent multiscale framework to simulate dynamic shock evolution over long runtimes and opens to the door to more complex studies involving shock propagation through composites and high-entropy alloys.
        \end{abstract}
    
    \end{frontmatter}
    
    \section{Introduction} \label{Sec: Introduction}
        Shock waves are complex events which can induce catastrophic damage to materials through plastic deformation and spall fracture.
        As such, considerable effort has been devoted towards understanding shock propagation within solids at the continuum level \cite{meyers1994dynamic,davison2008fundamentals}.
        However, a material's response to shock wave loading is linked to intricate behavior at the microscale.
        For example, fracture caused by a shock wave impact is the direct result of dislocations and void nucleation within the material's microstructure \cite{RustyGray2012,Fensin2014,Bingert2014}.
        Hence, it is imperative to understand shock wave propagation and evolution at the microscale in order to adequately predict material behavior at the macroscale.
        
        Atomistic shock wave simulations have been performed over the past several decades primarily using a technique known as non-equilibrium Molecular Dynamics (NEMD).
        In these simulations, the shock is typically generated by an impact or with a moving piston and is then allowed to propagate through the domain \cite{holian1995atomistic}.
        Recently, NEMD frameworks have been expanded to incorporate hundreds of millions of particles and have been used to model events such as dislocation generation \cite{germann2004dislocation,Tramontina2017Simulation,righi2021towards,zhu2021collapse}, twinning \cite{Higginbotham2013Molecular,wu2021unveiling,zhu2021novel}, void nucleation \cite{bringa2010void,bisht2019investigation,tian2021anisotropic}, and shock-induced spallation \cite{Srinivasan2007,Fensin2014Effect,wang2021spall,chen2021molecular,dewapriya2021molecular}.
        Unfortunately, NEMD techniques suffer from issues related to limited domain sizes and a large computational overhead which can cause artificial wave reflection and drastically restrict the total runtime \cite{kedharnath2021classical}.
        In the past two decades, alternative atomistic techniques have been developed to counteract such issues, and some examples are the uniaxial Hugoniostat \cite{maillet2000uniaxial,maillet2002uniaxial}, the multiscale shock technique (MSST) \cite{Reed2003,reed2006analysis}, and the moving window method \cite{Zhakhovskii1997,zhakhovsky2011two,davis2020one}.
        
        While modern atomistic techniques have greatly expanded our knowledge of shock wave behavior at the microscale, they nevertheless fail to capture the continuum-level response because the total number of particles that can be realistically incorporated into the domain is restricted by computer architecture and limited computational resources.
        To overcome these issues, \textit{concurrent multiscale} frameworks have been developed which retain atomistic information around a small region of interest and populate the remainder of the domain with finite elements \cite{kohlhoff1991crack,mcdowell2020connecting,van2020roadmap,xiong2021multiscale,fish2021mesoscopic}.
        A primary concern of concurrent methods is ensuring numerical compatibility at the atomistic-continuum (A-C) interfaces in order to reduce spurious wave reflections and ghost forces.
        Many schemes have been developed which address this issue in different ways \cite{tadmor2011modeling}, and a few of them are as follows: the Coupling of Length Scales (CLS) method \cite{rudd1998coarse}, the Bridging Domain (BD) method \cite{xiao2004bridging}, the Coupled Atomistic Discrete Dislocation (CADD) method \cite{Shilkrot2002Coupled}, and the Quasicontinuum (QC) method \cite{tadmor1996quasicontinuum}.
        Although methods such as these have had great success in material modeling, many of them still suffer from interface discrepancy due to a difference in governing equations between the atomistic and continuum regions.
        
        The Concurrent Atomistic-Continuum (CAC) method overcomes many of the A-C interface issues seen in other concurrent schemes by utilizing a unified multiscale framework built upon Atomistic Field Theory \cite{chen2005atomistic,chen2009reformulation} whereby a single set of governing equations is employed throughout the entire domain \cite{xiong2011coarse,yang2013concurrent,xiong2014prediction,xiong2015concurrent,xu2016mesh,chen2017recent,chen2018passing,xu2018pycac,chen2019concurrent}.
        As a result, CAC has seen tremendous success over the past decade in modeling phenomena such as dislocations and grain boundaries \cite{xiong2014sub,chen2017effects} as well as passing high-frequency waves between the atomistic and continuum regions \cite{chen2018passing,DAVIS2022111702}.
        Recent work has even implemented an A-atom approach within CAC to perform large-scale simulations of multicomponent alloys \cite{chu2022multiscale}, and research of dislocation evolution \cite{selimov2021lattice} as well as crystal plasticity \cite{selimov2022coarse} is ongoing.
        Unfortunately, the study of shock wave propagation using the CAC method has been limited due to the highly dynamic nature of such phenomena.
        While previous work has addressed this complication by incorporating moving window techniques into a CAC framework to track a nonlinear shock wave for long runtimes \cite{davis2022moving}, this formulation only considered a 1D chain of particles and was thus limited in scope.
        
        In the present work, we develop a multiscale framework using the CAC method to model long-time shock wave propagation through a two-dimensional lattice.
        Specifically, we utilize both the Hugoniot shock equations \cite{meyers1994dynamic} as well as the nonlinear Eulerian thermoelastic shock equations \cite{clayton2013nonlinear} to study the classic Riemann problem of a single traveling discontinuity. 
        Furthermore, we enhance the moving window techniques first presented in \cite{davis2022moving} to track the shock over long simulation times and engineering-scale domains.
        Each method maintains the shock front at the center of the atomistic region for the entire runtime, so the wave front never encounters the A-C interfaces.
        This allows us to model shock propagation for greater simulation times than traditional NEMD and multiscale methods and thus gain valuable information about the long-term, time-averaged material response to shock loading of two different FCC solids.
        
        This paper is organized as follows.
        Section \ref{Sec: Shock Wave Background} characterizes the shocks studied in the present work and elaborates on both the Hugoniot and Eulerian analytical models.
        Section \ref{Sec: Computational Framework} describes the framework's geometry and boundary conditions as well as presents the interatomic potential, thermostat, material parameters, and shock constants utilized in the simulations.
        Section \ref{Sec: CAC Method} discusses the finite element formulation of CAC and its 2D implementation.
        Section \ref{Sec: Shock Propagation Technique} outlines both the shock propagation technique and the two moving window schemes used to track the shock front.
        Section \ref{Sec: Elastic Anisotropy: Crystal Orientation Dependence on Shock Propagation Response} presents shock propagation results obtained with the conveyor technique and compares these to both analytical models to highlight the directional anisotropies in single crystals subject to shock loading.
        Section \ref{Sec: Results with the Coarsen-Refine Method and Formulation Efficiency} uses the coarsen-refine technique to perform parametric studies related to the shock front's structure as well as showcases the efficiency of the current model compared to NEMD simulations.
        Finally, Section \ref{Sec: Conclusion} concludes the paper and discusses ideas for future work.
        
    \section{Shock Wave Background} \label{Sec: Shock Wave Background}
    
        \subsection{Problem statement} \label{Sec: Problem statement}
\begin{figure}[htpb]
                \centering
                \includegraphics[width=0.45\textwidth]{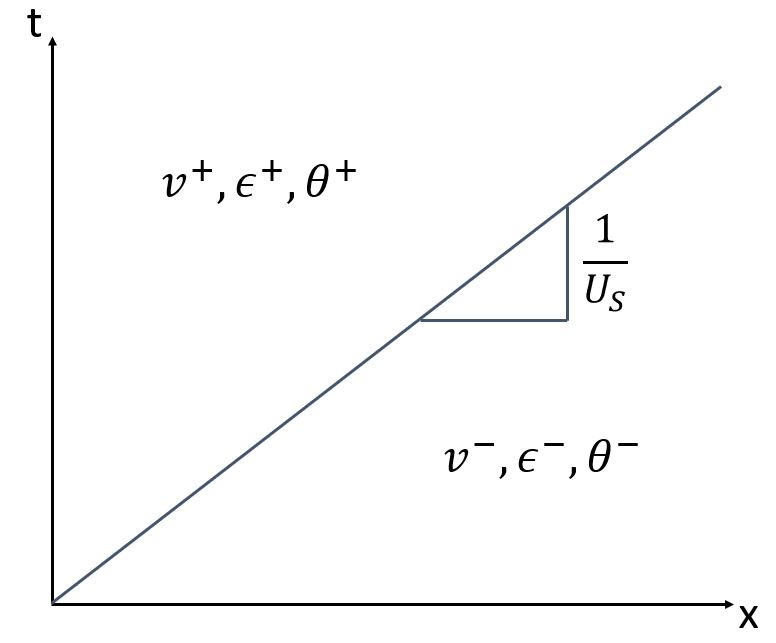}
                \caption{Riemann problem of a shock wave with constant states in front of and behind the shock front.}
                \label{Fig:Riemann Shock}
            \end{figure}

            We consider a two-dimensional monatomic lattice with no defects under compression by an ideal longitudinal shock wave traveling in the $x-$direction.
            Mathematically, we represent the shock as a propagating discontinuity across which there exists a jump in particle velocity ($v$), stress ($\sigma$), strain ($\epsilon$), and temperature ($\theta$).
            Material quantities ahead of the shock front have the superscript \textit{$-$}, and quantities behind the shock front have the superscript \textit{$+$}.
            The notation $\llbracket\cdot\rrbracket$ denotes the change in a given quantity $(\cdot)$ across the shock front.
            During each simulation, particles ahead of the shock wave are assumed to be at zero mean particle velocity, unstressed, unstrained, and at room temperature ($295$ K).
            Furthermore, the shock propagates at a natural velocity $U_S$ along the surface of the primitive unit cell of an FCC lattice. 
            We incorporate these parameters into a moving window CAC framework to simulate long-time shock wave propagation over engineering-scale domains.
            Specifically, we model the classic Riemann problem of a single shock wave front with constant states on either side as shown in Fig. \ref{Fig:Riemann Shock}.
        
            To calculate the aforementioned jump parameters and thus characterize the shock wave at the continuum level, we use two different formulations which are discussed below.
    
        \subsection{Hugoniot shock equations} \label{Sec: Hugoniot shock equations}
            First, we simulate dynamic shock wave propagation and evolution using the conservation of momentum, continuity equation, Hugoniot equation of state (EOS), and a thermodynamic relationship derived from the shock Hugoniot and release isentrope.
            By applying the conservation of linear momentum and continuity of displacement across the discontinuous shock front and assuming uniaxial loading, we obtain the following standard one-dimensional shock wave jump equations \cite{davison2008fundamentals}:
            \begin{align}
                \llbracket \sigma \rrbracket + \rho U_s \llbracket v \rrbracket &= 0 \label{Eq: Momentum Equation} \\
                \llbracket v \rrbracket + U_s \llbracket \epsilon \rrbracket &= 0\label{Eq: Continuity Equation}
            \end{align}
            where $\rho$ denotes the density of the material.
            To fully parameterize the system, Eqs. (\ref{Eq: Momentum Equation}) and (\ref{Eq: Continuity Equation}) are supplemented by an empirically observed linear relation between shock velocity and particle velocity \cite{meyers1994dynamic}:
            \begin{equation}
                U_s = C_0 + S\llbracket v\rrbracket. \label{Eq: Linear Law}
            \end{equation}
            Here, \textit{S} is a dimensionless, empirical parameter representing the slope of the shock velocity vs. particle velocity Hugoniot curve, and \textit{$C_{0}$} is the sound velocity in the material at zero stress.
            We can use Eqs. (\ref{Eq: Momentum Equation}), (\ref{Eq: Continuity Equation}), and (\ref{Eq: Linear Law}) to derive the standard Hugoniot stress-strain relationship given as follows: 
            \begin{equation}
                \sigma = \frac{\rho C_0^2 \llbracket \epsilon \rrbracket}{(1 + S \llbracket \epsilon \rrbracket)^2} \label{Eq: Hugoniot EOS}
            \end{equation}
            where compressive stress and strain are considered positive.
            The Hugoniot stress-strain relationship forms the basis of modern equations of state.
            Finally, we calculate the rise in temperature across the shock front by solving the following ordinary differential equation \cite{davison2008fundamentals}:
            \begin{equation}
                C_{V} \left(\frac{d\theta}{d \epsilon} \right)_{H} - \frac{\gamma \theta C_{V}}{1 - \epsilon} = \frac{\epsilon}{2} \left(\frac{d \sigma}{d \epsilon} \right)_{H} - \frac{\sigma}{2} \label{Eq: Shock Heat Equation}
            \end{equation}
            where $C_V$ is the volumetric specific heat capacity, and $\gamma$ is the Mie-Gruneisen parameter for the material. 
        
        \subsection{Eulerian shock equations} \label{Sec: Eulerian shock equations}
            We also characterize the propagating shock wave using the nonlinear Eulerian thermoelastic shock equations derived in \cite{clayton2013nonlinear,clayton2014shock} for anisotropic crystals.
            Nonlinear elastic constitutive models of material behavior which do not account for slippage and plasticity are generally idealizations because even small uniaxial compressive strains can cause ductile materials to reach the experimental Hugoniot elastic limit (HEL).
            However, such elastic formulations can be practically applied to defect-free atomistic and multiscale simulations since these domains may be shocked to finite strains over relatively short time scales and small volumes \cite{clayton2014shock,zimmerman2011elastic}.
            
            For an extensive derivation of the Eulerian formulation for shock waves, we refer the reader to \cite{clayton2013nonlinear}. 
            Here, we merely present the relevant equations used in the current work.
            The particle velocity in the shocked material, shock propagation velocity, and temperature in the shocked material are given by the following respective equations:
            \begin{equation} \label{Eq: ClaytonEulerianParticleVelocity}
                v = \biggl \{ \left(\frac{\hat{S}}{\rho} \right) \left[ \left(1-2D \right) - \left(1-2D \right)^{3/2} \right] \biggr \}^{1/2}
            \end{equation}
            \begin{equation} \label{Eq: ClaytonEulerianShockVelocity}
                U_S = v \left[1 - \left(1-2D \right)^{-1/2} \right]^{-1}
            \end{equation}
            \begin{equation} \label{Eq: ClaytonEulerianTemperature}
                \theta = \frac{\partial \hat{U}}{\partial \eta} = \theta_0 \left(1 - \Hat{\Gamma}_1 D - \frac{1}{2} \hat{\Gamma}_{11} D^2 \right).
            \end{equation}
            In Eqs. (\ref{Eq: ClaytonEulerianParticleVelocity}), (\ref{Eq: ClaytonEulerianShockVelocity}), and (\ref{Eq: ClaytonEulerianTemperature}), $D$ is the Eulerian strain represented by the following expression:
            \begin{equation} \label{Eq: EulerianStrain}
                D = \frac{1}{2} \left(1 - F^{-2} \right) = \frac{1}{2} \left[1 - \frac{1}{(1 + \epsilon)^2} \right].
            \end{equation}
            Here, the term \textit{Eulerian} refers to a strain which is a function of the inverse deformation gradient $F$.
            Hence, the strain tensor $D$ assumes material coordinates rather than spatial coordinates, so it can be applied to simulations of anisotropic materials \cite{clayton2013nonlinear}. 
            Furthermore, $\hat{U}$ is the Eulerian fourth-order internal energy function as seen below \cite{clayton2014shock}:
            \begin{equation} \label{Eq: ClaytonEulerianEnergy}
                \Hat{U} = \frac{1}{2}\Hat{C}_{11}D^2 + \frac{1}{6}\Hat{C}_{111}D^3 + \frac{1}{24} \Hat{C}_{1111}D^4 - \theta_0 \left(\Hat{\Gamma}_{1} D + \frac{1}{2} \Hat{\Gamma}_{11}D^2 - 1 \right)\eta
            \end{equation}
            where $\Hat{C}_{11}$, $\Hat{C}_{111}$, and $\Hat{C}_{1111}$ are the Eulerian second, third, and fourth-order elastic constants, $\Hat{\Gamma}_{1}$ and $\Hat{\Gamma}_{11}$ are the Eulerian first and second-order Gr\"{u}neisen parameters, and $\eta = 0$ is the entropy ahead of the shock front.
            The elastic constants and Gr\"{u}neisen parameters in an Eulerian setting are obtained from their non-Eulerian counterparts using the following relations \cite{weaver1976application,perrin1978application,clayton2013nonlinear}:
            \begin{align}
                \Hat{C}_{11} &= C_{11} \label{Eq: ClaytonEulerianC11} \\
                \Hat{C}_{111} &= C_{111} + 12C_{11} \label{Eq: ClaytonEulerianC111} \\
                \Hat{C}_{1111} &= C_{1111} - 18C_{111} - 318C_{11} \label{Eq: ClaytonEulerianC1111} \\
                \Hat{\Gamma}_{1} &= \Gamma_{1} \\
                \Hat{\Gamma}_{11} &= \Gamma_{11} + 4\Gamma_1 \label{Eq: ClaytonEulerianGruneisenConstant}.
            \end{align}
            Finally, the conjugate stress $\Hat{S}$ is represented by 
            \begin{align} \label{Eq: ClaytonEulerianStress}
                \Hat{S} &= \frac{\partial \Hat{U}}{\partial D} \\ \nonumber
                &= C_{11}D + \frac{1}{2}\hat{C}_{111}D^2 + \left(\frac{1}{6}\hat{C}_{1111} - \theta_0 \Hat{\Gamma}_1 b_3 \right)D^3 - \theta_0 D^4 \left[ \left(\Hat{\Gamma}_1 b_4 + \hat{\Gamma}_{11} b_3 \right) + \left(\Hat{\Gamma}_1 b_5 + \hat{\Gamma} _{11} b_4 \right) D \right]
            \end{align}
            where $b_3$, $b_4$, and $b_5$ are polynomials for entropy $\eta$ generated across the shock front, and their expressions can be found in \cite{clayton2013nonlinear}.
            In each shock wave simulation, we use the fourth-order expression of Eqs. (\ref{Eq: ClaytonEulerianParticleVelocity}), (\ref{Eq: ClaytonEulerianShockVelocity}), and (\ref{Eq: ClaytonEulerianTemperature}).
            
        \subsection{`Elastic' shock waves} \label{Sec: Elastic shock waves}
            To legitimately utilize the shock equations from Sec. \ref{Sec: Eulerian shock equations} as well as avoid intractability with the moving window techniques, we perform shock simulations with relatively small strains such that the resulting stresses are below the HEL of the material (see \ref{App: Stress-strain relations}). 
            To maintain consistency, we refer to these as \textit{elastic} shock waves in the present work, but they are also classified as \textit{weak} shocks in other papers \cite{holian1995atomistic}.
            Elastic shock waves are often modeled in defect-free crystals with NEMD techniques to study a particular phenomenon, test a new framework, or validate a given potential \cite{holian1978molecular,zimmerman2011elastic,davis2020one}, and their distinguishing characteristic is the lack of any permanent dislocations (inelastic deformation) behind the wave front.
            This is possible because the HEL is typically higher than what is seen in experimental settings \cite{yang2013concurrent}, and the wave speed is still greater than the sound velocity in the material at the low strains.
            Modeling shock propagation with the CAC moving window framework using thermoelastic-viscoplastic models \cite{lloyd2014simulation,lloyd2014plane} is a worthy pursuit but would add an extra layer of complexity to the current model and is thus reserved for future studies.
       
    \section{Computational Framework} \label{Sec: Computational Framework}
    
        \subsection{Geometry and boundary conditions} \label{Sec: Geometry and boundary conditions}
\begin{figure}[htpb]
                \centering
                \includegraphics[width=0.9\textwidth]{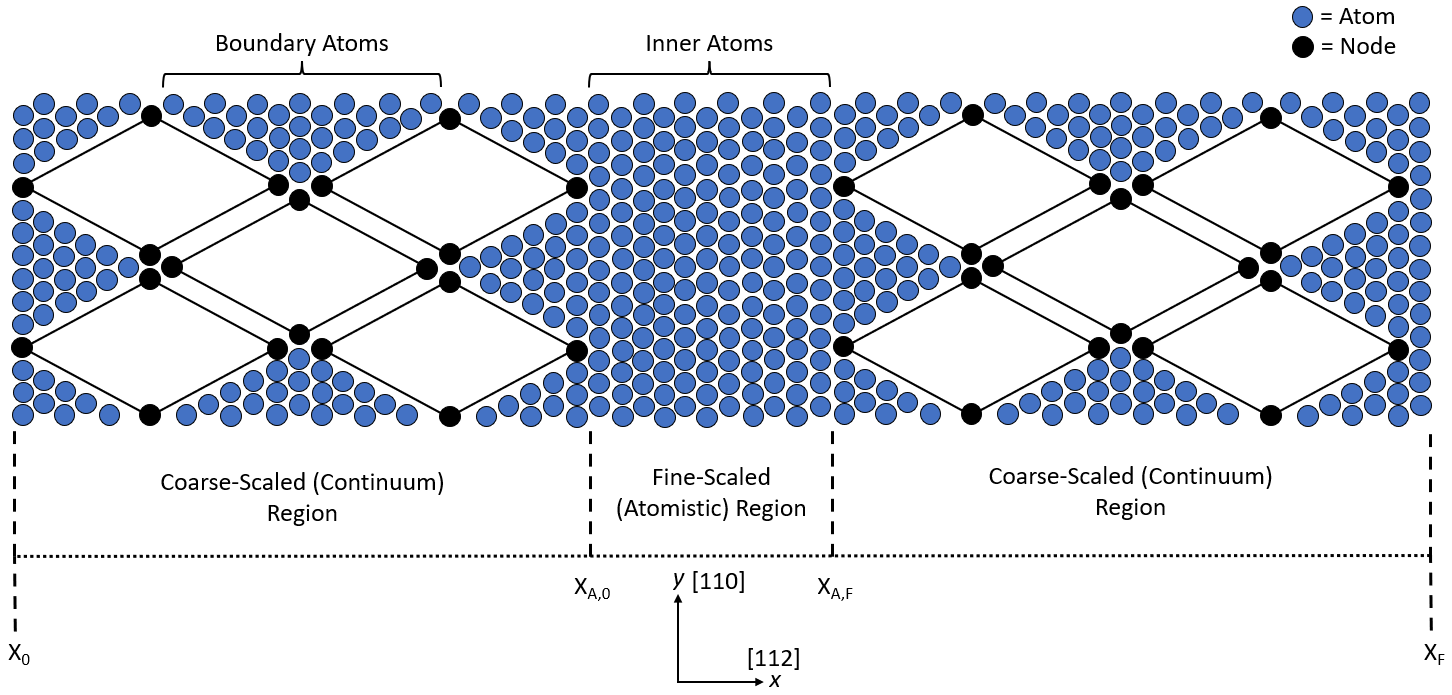}
                \caption{Schematic of the two-dimensional CAC framework.}
                \label{Fig:2D_CAC_Framework}
            \end{figure}

            The two-dimensional CAC framework is implemented using an in-house C++ code, and the monatomic lattice is split into three primary regions as seen in Fig. \ref{Fig:2D_CAC_Framework}.
            The two coarse-scaled (continuum) regions are composed of rhombus elements, and the four particles which make up any particular element are referred to as \textit{nodes} in the present work.
            We choose rhombus elements because they align with the primitive unit cell of the FCC lattice (see Sec. \ref{Sec: Two-dimensional formulation}) and thus facilitate a smooth transition between the fine-scaled and coarse-scaled regions.
            Specifically, the $x$-direction corresponds to the [112] lattice orientation while the $y$-direction corresponds to the [110] lattice orientation.
            Since element connectivity is not required in CAC \cite{xiong2011coarse}, each node is a member of only one element, and this greatly reduces the complexity of the finite element formulation.
            Furthermore, the edges of the grid in the continuum regions are ``filled in" with particles which we refer to as \textit{boundary atoms} in this work.
            This is done in order to facilitate periodic boundary conditions as shown in \cite{xu2015quasistatic}.
            
            The two coarse-scaled regions flank the inner fine-scaled (atomistic) region on the left and right-hand side, and we refer to the particles in this region as either \textit{inner atoms} or just \textit{atoms} in the present work.
            The ``elements" in the fine-scaled region are reduced to their smallest possible configuration such that only four atoms constitute the entire area of each element.
            Hence, both the fine-scaled and coarse-scaled regions are technically made up of rhombuses with the only differences being the area and mass of their respective elements.
            As a consequence, one governing equation along with a single mass matrix is utilized for both regions, all force calculations are nonlocal, and the interatomic potential is the only constitutive relation \cite{xu2016mesh}.
            Thus, the particles at the A-C interfaces ($x_{A,0}$ and $x_{A,F}$) interact with each other directly without creating ghost forces \cite{xu2015quasistatic,xu2018pycac}.
            
            We note that to avoid introducing non-physical strains into the domain during shock simulations, semi-periodic boundary conditions are employed in the $x$-direction whereby the particles at the ends of the chain ($x_0$ and $x_F$) are made neighbors with the nodes at the interfaces ($x_{A,0}$ and $x_{A,F}$ respectively) \cite{davis2022moving}. 
            Additionally, since the present work only considers uniaxial compression, we utilize periodic boundary conditions in the $y$-direction when modeling a longitudinal shock wave.
  
        \subsection{Interatomic potential and material parameters} \label{Sec: Interatomic potential and material parameters}
            To calculate the integrand of the internal force density (Eq. \ref{Eq: Internal Force Density}), we use the modified Morse interatomic potential function.
            This potential only considers first nearest neighbor interactions and is given by the following expression \cite{macdonald1981thermodynamic}:
            \begin{equation} \label{Eq: Morse}
                \Pi(r_{ij}) = \frac{D_0}{2B-1}\left[\mathrm{e}^{-2 \alpha \sqrt{B} (r_{ij} - r_0)} - 2B\mathrm{e}^{-\alpha (r_{ij} - r_0) / \sqrt{B}}\right]
            \end{equation}
            where $r_{ij} = |\textbf{x}_{i} - \textbf{x}_{j}|$ is the magnitude of the displacement between particle $i$ and $j$, and $r_0$ is the distance at which the potential reaches the minimum (defined as the close-packed neighbor spacing).
            We perform shock simulations with Cu and Al, and the parameters for these materials are given in Table \ref{Table: MorsePotentialParameters}.
            \begin{table}[h]
                \centering
                \caption{Material constants and Morse parameters of two different FCC metals \cite{macdonald1981thermodynamic}.}
                \label{Table: MorsePotentialParameters}
                \begin{tabular}{||c  c  c  c  c  c  c  c||}
                \hline
                \textit{Element} & \textit{mass (u)} & \textit{$\rho_0$ (g/$cm^3$)} & 
                $\Gamma_1$ & \textit{$r_0$ (\AA)} & \textit{$a$ (\AA$^{-1}$)} & \textit{$D_0$ (eV)} & \textit{B} \\
                \hline
                Cu & 63.55 & 8.96 & 1.97 & 2.5471 & 1.1857 & 0.5869 & 2.265 \\
                Al & 26.98 & 2.70 & 2.17 & 2.8485 & 1.1611 & 0.3976 & 2.5 \\
                \hline 
                \end{tabular}
            \end{table}
            
        \subsection{Integration algorithm and thermostat} \label{Sec: Integration algorithm and thermostat}
            The CAC governing equation (Eq. \ref{Eq: Matrix Form of Governing Equation}) is a second-order ordinary differential equation in time, and we solve it using the velocity Verlet algorithm.
            The time step used in the integration algorithm is chosen to be $\Delta t = 0.001$ ps in order to minimize numerical error.
            
            To apply temperature to the domain, we use the Langevin thermostat -- a stochastic thermostat which adds a random force to the particle motion along with a damping term $\zeta$.  
            In particular, we modify the velocity Verlet algorithm in the presence of the Langevin thermostat by performing the discretization used in LAMMPS \cite{schneider1978molecular}:
            \begin{align}
                \textbf{v}_{i} \left(t + \frac{\Delta t}{2} \right) &= \textbf{v}_{i}(t) - \frac{\Delta t}{2} \left[\frac{\nabla_{i} \Pi(t)}{m} + \zeta \textbf{v}_{i}(t) \right] + \sqrt{\frac{\Delta t k_B \theta \zeta}{m}}\tilde{\textbf{h}}_{i}
                \nonumber \\
                \textbf{x}_{i}(t + \Delta t) &= \textbf{x}_{i}(t) + \textbf{v}_{i} \left(t + \frac{\Delta t}{2} \right) \Delta t
                \nonumber \\
                \textbf{v}_{i} \left(t + \Delta t \right) &= \textbf{v}_{i} \left(t + \frac{\Delta t}{2} \right) - \frac{\Delta t}{2} \left[\frac{\nabla_{i} \Pi(t + \Delta t)}{m} + \zeta \textbf{v}_{i} \left(t + \frac{\Delta t}{2} \right) \right] + \sqrt{\frac{\Delta t k_B \theta \zeta}{m}}\tilde{\textbf{h}}_{i}.
            \end{align}
            Here, $\textbf{x}_{i}$ and $\textbf{v}_{i}$ denote the position and velocity of the ${i}^{th}$ particle, $m$ is the atomic mass, $k_{B}$ is Boltzmann's constant, and $\tilde{\textbf{h}}_{i}$ is a Gaussian random variable with a mean of zero and a variance of one. 
            As per Langevin's requirements, we generate a different random variable for each particle during each velocity update.
            Since Langevin is local in nature, the target temperatures $\theta^+$ and $\theta^-$ are specified for each particle. 
            For the compressive strains applied in this work, $\theta^+$ has an upper boundary of $\sim$ 450 K.
            
        \subsection{Shock parameters} \label{Sec: Shock parameters}
            In Table \ref{Table: ShockParameters}, we present the empirical Hugoniot shock parameters as well as the second, third, and fourth-order elastic constants (in a normal and Eulerian setting) for both Cu and Al.
            The Hugoniot parameters are obtained from \cite{marsh1980lasl}, the second and third-order elastic constants for Cu and Al are obtained from \cite{hiki1966anharmonicity} and \cite{thomas1968third} respectively, and the fourth-order elastic constants are obtained from \cite{clayton2014shock}.
            For these values, the temperature is assumed to be $295$ K, $C_0$ is given in $km/sec$, $S$ is unitless, and the elastic constants are given in GPa.
            The Hugoniot parameters are derived for a shock wave propagating through a bulk, polycrystalline material.
            Furthermore, the elastic constants represent the pure-mode directions such that a planar shock impact results in an exclusively longitudinal component (along the [100] direction) with no transmitted shear stress, and hence the one-dimensional analysis is valid.
            We use these parameters as initial input in our shock simulations and compare the results from the CAC model to analytical and empirical data in Sec. \ref{Sec: Elastic Anisotropy: Crystal Orientation Dependence on Shock Propagation Response}.
            \begin{table}[h]
                \centering
                \caption{Hugoniot and Eulerian shock parameters for Cu and Al ($\theta = 295$ K, $C_0$ in km/sec, and $C_{\alpha \beta}$ in GPa).}
                \label{Table: ShockParameters}
                \begin{tabular}{||c  c  c||}
                \hline
                \textit{Property} & \textit{Cu [100]} & \textit{Al [100]} \\
                \hline
                $C_0$ & 3.94 & 5.33 \\
                S & 1.49 & 1.34 \\
                $C_{11}$ & 166 & 107 \\
                $C_{111}$ & -1270 & -1080 \\
                $\Hat{C}_{111}$ & 722 & 204 \\
                $C_{1111}$ & 11900 & 25000 \\
                $\Hat{C}_{1111}$ & 2000 & 10500 \\
                \hline 
                \end{tabular}
            \end{table}
            
    \section{CAC Method} \label{Sec: CAC Method}
    
        \subsection{Finite element implementation} \label{Sec: Finite element implementation}
            Here, we give a very brief overview of the finite element implementation of CAC, but more details can be found in \cite{xiong2009multiscale,deng2010coarse,xiong2011coarse,chen2019concurrent}.
            The mathematical foundation of CAC is Atomistic Field Theory (AFT), and the governing equations of AFT have a similar form to the balance laws of classical continuum mechanics. 
            %
            Exploiting the definitions of internal force density and kinetic temperature derived in \cite{chen2005nanoscale} and \cite{chen2006local}, we can recast the instantaneous balance equation of linear momentum as follows \cite{xiong2009multiscale}:
            \begin{equation} \label{Eq: FEM Governing Equation 2} 
                \rho^{\alpha} \Ddot{\textbf{u}}^{\alpha}(\textbf{x}) = \textbf{f}_{int}^{\alpha}(\textbf{x}) + \textbf{f}^{\alpha}(\textbf{x}).
            \end{equation}
            In Eq. (\ref{Eq: FEM Governing Equation 2}), $\textbf{u}^{\alpha}(\textbf{x})$ is the displacement of the $\alpha^{th}$ atom in the unit cell, $\rho^{\alpha} = m^\alpha/\Delta V $ is the volumetric mass density, $m^{\alpha}$ is the mass of the $\alpha^{th}$ atom, $\Delta V$ is the volume of the unit cell, $\textbf{f}_{int}^{\alpha}(\textbf{x})$ is the internal force density, and $\textbf{f}^{\alpha}(\textbf{x})$ is the force density due to external forces and temperature.
            The two terms on right side of Eq. (\ref{Eq: FEM Governing Equation 2}) are represented as follows:
            \begin{align} 
                \textbf{f}_{int}^{\alpha}(\textbf{x}) &= \int_{\Omega(\textbf{x}')} \sum_{\beta = 1}^{N_a} \textbf{f} \left[\textbf{u}^{\alpha}(\textbf{x}) - \textbf{u}^{\beta}(\textbf{x}') \right] d\textbf{x}' \label{Eq: Internal Force Density} \\ 
                \textbf{f}^{\alpha}(\textbf{x}) &= \textbf{f}_{ext}^{\alpha}(\textbf{x}) - \frac{m^{\alpha} k_B}{M \Delta V} \nabla_{\textbf{x}} \theta^{\alpha} \label{Eq: External Force Density}
            \end{align}
            where $\textbf{f}_{ext}^{\alpha}(\textbf{x})$ is the external force density, $M$ is the total mass of the atoms within a unit cell, and $\theta^{\alpha}$ is the kinetic temperature.
            We note that the internal force density can be obtained exclusively from the interatomic potential function since it is a nonlinear, nonlocal function of relative displacements between neighboring particles \cite{yang2014concurrent}. 
        
            We employ the finite element method to calculate the numerical solution of Eq. (\ref{Eq: FEM Governing Equation 2}).
            We populate the domain with finite elements such that every element contains a collection of primitive unit cells. 
            Each nodal location represents a unit cell which is itself made up of particles.
            As a result, CAC provides a two-level description of crystals and follows the solid state physics model whereby the structure is continuous at the lattice level but discrete at the atomic scale.
            We use interpolation within each element in the domain to approximate the displacement field as follows \cite{xiong2011coarse}:
            \begin{equation} \label{Eq: Approximate Displacement Field}
                \Hat{\textbf{u}}^{\alpha}(\textbf{x}) = \boldsymbol{\Phi}_{\xi}(\textbf{x}) \textbf{U}_{\xi}^{\alpha}.
            \end{equation}
            In Eq. (\ref{Eq: Approximate Displacement Field}), $\Hat{\textbf{u}}^{\alpha}(\textbf{x})$ is the displacement field for the $\alpha^{th}$ atom within a given element, $\boldsymbol{\Phi}_{\xi}(\textbf{x})$ is the shape function, and $\textbf{U}_{\xi}^{\alpha}$ is the displacement of the $\alpha^{th}$ atom within the $\xi^{th}$ element node. 
            We let $\xi = 1, 2, ..., n$ where $n$ is the total number of nodes in the element (four in this work).
            
            Applying the method of weighted residuals, we obtain the weak form of the governing equation by multiplying Eq. (\ref{Eq: FEM Governing Equation 2}) with a weight function $\boldsymbol{\Phi}_{\eta}(\textbf{x})$ and integrating over the entire domain:
            \begin{equation} \label{Eq: Weak Form of Governing Equation 1}
                \int_{\Omega(\textbf{x})} \left[\rho^{\alpha} \boldsymbol{\Phi}_{\eta}(\textbf{x})  \Ddot{\textbf{u}}^{\alpha}(\textbf{x}) \right] d\textbf{x} = \int_{\Omega(\textbf{x})} \left[ \boldsymbol{\Phi}_{\eta}(\textbf{x}) \textbf{f}_{int}^{\alpha}(\textbf{x}) \right] d\textbf{x} + \int_{\Omega(\textbf{x})} \left[ \boldsymbol{\Phi}_{\eta}(\textbf{x}) \textbf{f}^{\alpha}(\textbf{x}) \right] d\textbf{x}.
            \end{equation}
            Specifically, the Galerkin method is used to obtain the above expression, so the weight function $\boldsymbol{\Phi}_{\eta}(\textbf{x})$ equals the shape function $\boldsymbol{\Phi}_{\xi}(\textbf{x})$ in this case.
            Substituting Eqs. (\ref{Eq: Internal Force Density}), (\ref{Eq: External Force Density}), and (\ref{Eq: Approximate Displacement Field}) into Eq. (\ref{Eq: Weak Form of Governing Equation 1}), we arrive at the weak form of the CAC governing equation which can be represented in matrix form as follow:
            \begin{equation} \label{Eq: Matrix Form of Governing Equation}
                \textbf{M}^{\alpha} \Ddot{\textbf{U}}_{\xi}^{\alpha} = \textbf{F}_{int}^{\alpha} + \textbf{F}^{\alpha}
            \end{equation}
            where
            \begin{align}
                \textbf{M}^{\alpha} &= \int_{\Omega(\textbf{x})} \left[\rho^{\alpha} \boldsymbol{\Phi}_{\eta}(\textbf{x}) \boldsymbol{\Phi}_{\xi}(\textbf{x}) \right] d\textbf{x} \label{Eq: Inertial term} \\
                \textbf{F}_{int}^{\alpha} &= \int_{\Omega(\textbf{x})} \boldsymbol{\Phi}_{\eta}(\textbf{x}) \int_{\Omega(\textbf{x}')} \sum_{\beta = 1}^{N_a} \textbf{f} \left[\boldsymbol{\Phi}_{\xi}(\textbf{x}) \textbf{U}_{\xi}^{\alpha} - \boldsymbol{\Phi}_{\xi}(\textbf{x}') \textbf{U}_{\xi}^{\beta} \right] d\textbf{x}' d\textbf{x} \\
                \textbf{F}^{\alpha} &= \int_{\Omega(\textbf{x})} \left[ \boldsymbol{\Phi}_{\eta}(\textbf{x}) \textbf{f}^{\alpha}(\textbf{x}) \right] d\textbf{x}.
            \end{align}
            In this work, we approximate the inertial term (Eq. \ref{Eq: Inertial term}) using the lumped mass matrix derived in \ref{App: Mass matrix}.
            Additionally, no external forces are applied, and temperature is incorporated via a thermostat as in \cite{xiong2014prediction} and \cite{chen2018passing}. 
            The internal force density $\textbf{F}_{int}^{\alpha}$ is the most computationally expensive term, and we evaluate it numerically using Gaussian integration as discussed in \ref{App: Gaussian integration}. 
            
            By using this finite element implementation of CAC, a majority of the degrees of freedom in the coarse-scaled regions are eliminated. 
            For critical regions where atomistic behavior is required, the finest mesh is used such that each rhombus ``element" consists exclusively of four atoms with no additional lattice points.
            Thus, CAC uses AFT to produce a unified theoretical framework between the fine-scaled and coarse-scaled regions.
            A unique feature of CAC is that in the finite element implementation, element connectivity is not required because the nonlocal interatomic force field is the only constitutive relation \cite{xiong2011coarse}.
            This is similar to aspects of the cohesive zone model \cite{needleman1987continuum} and greatly simplifies the implementation of both the mass matrix as well as the force calculations.
    
        \subsection{Two-dimensional formulation} \label{Sec: Two-dimensional formulation}
\begin{figure}[htpb]
                \centering
                \begin{subfigure}{0.35\textwidth}
                    \centering
                    \includegraphics[width=\textwidth]{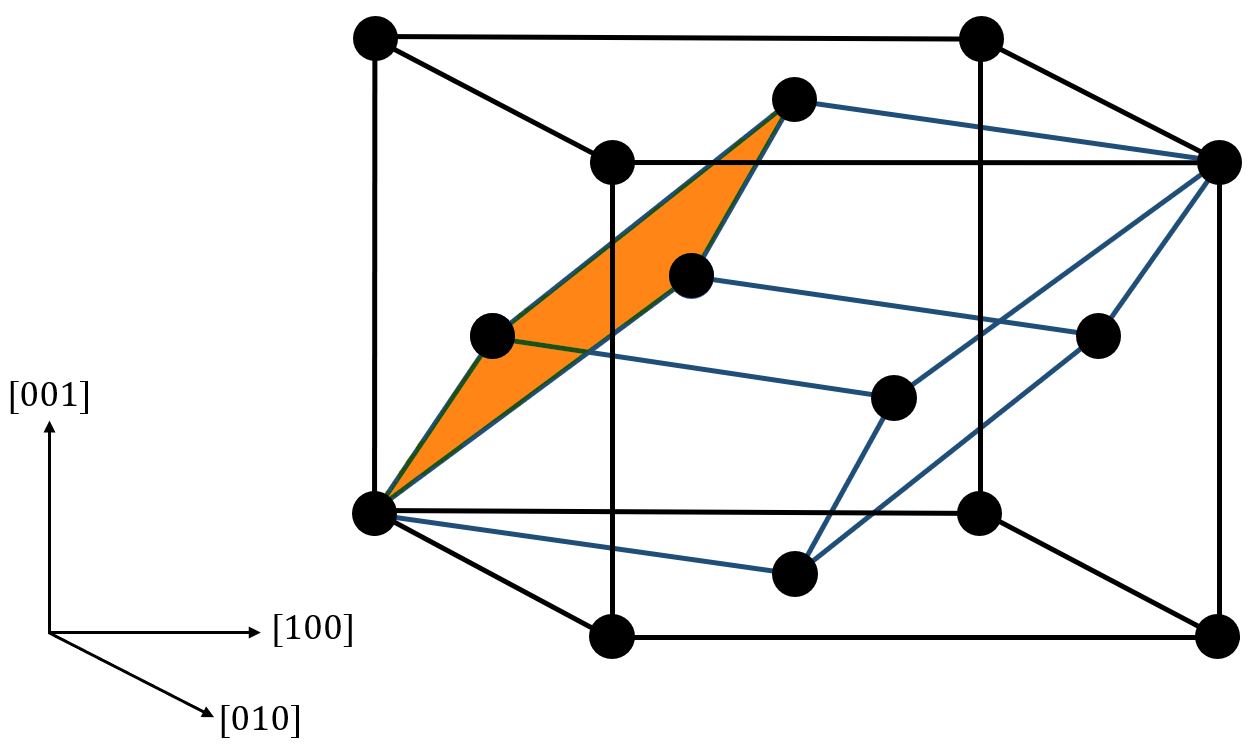}
                    \caption{}
                    \label{Fig:Primitive_Unit_Cell}
                \end{subfigure}
                \begin{subfigure}{0.61\textwidth}
                    \centering
                    \includegraphics[width=0.5\textwidth]{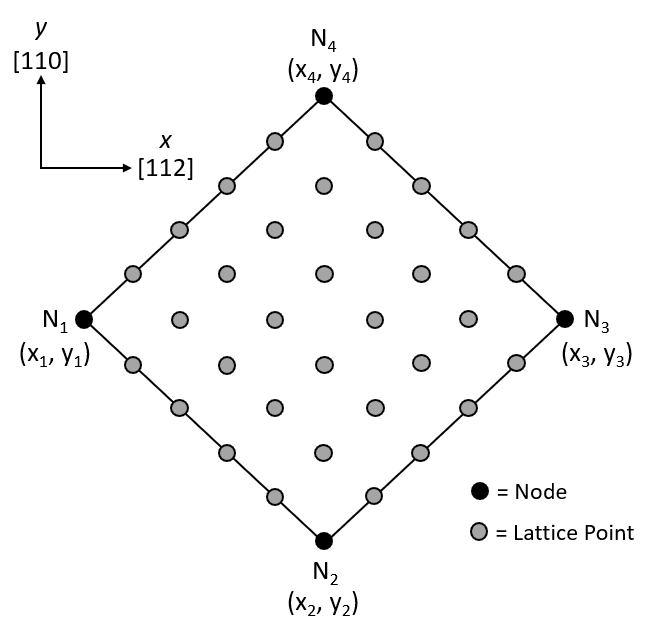}
                    \caption{}
                    \label{Fig:RhombusElementSchematic}
                \end{subfigure}
                \caption{(a) Rhombohedral element constituting the primitive unit cell (blue lines) of an FCC lattice. 
                The shaded region represents the two-dimensional rhombus element utilized in the present formulation.
                (b) Schematic of the two-dimensional rhombus element.}
            \end{figure}

            Rhombohedral elements are utilized within the CAC formulation to replicate the primitive unit cell of a monocrystalline lattice (FCC in the present work). 
            A sketch of this can be seen in Fig. \ref{Fig:Primitive_Unit_Cell}, where we observe the primitive unit cell (blue lines) within the broader FCC crystal structure.
            Furthermore, the shaded region represents the two-dimensional atomic plane used in our formulation whereby rhombus elements are incorporated throughout the domain.
            Since the same constitutive relation is used both within elements as well as between elements, dislocations and cracks emerge naturally through the separation of finite elements \cite{xiong2011coarse}.
            This is a direct result of the CAC governing equations, and it allows such defects to pass smoothly between the atomistic and continuum regions without deforming individual elements.
            
            A schematic of the two-dimensional rhombus element can be seen in Fig. \ref{Fig:RhombusElementSchematic}.
            Here, the black circles represent the four nodes where the governing equations are applied, and the grey circles represent the lattice points which serve as nodal neighbors and thus aid in the force calculations.
            For monatomic crystals, each nodal location (unit cell) only contains one atom, and the positions of the lattice points are interpolated using Eq. (\ref{Eq: Approximate Displacement Field}) throughout the element.
            We emphasize that the lattice points are excluded from the Verlet algorithm.
            Finally, since no external forces are applied in this work, the governing equations from Sec. \ref{Sec: Finite element implementation} reduce to the following:
            \begin{equation} \label{Eq: Matrix Form of Governing Equation in 2D}
                \textbf{M} \Ddot{\textbf{U}} - \textbf{F}^{int} = \textbf{0}
            \end{equation}
            where
            \begin{equation} \label{Eq: Mass Matrix in 2D}
                \textbf{M} = \int_{\Omega(\textbf{x})} \left[\rho \boldsymbol{\Phi}(\textbf{x}) \boldsymbol{\Phi}(\textbf{x}) \right] d\textbf{x}
            \end{equation}
            \begin{equation} \label{Eq: Internal Force Density in 2D}
                \textbf{F}^{int} = \int_{\Omega(\textbf{x})} \boldsymbol{\Phi}(\textbf{x}) \int_{\Omega(\textbf{x}')} \sum_{j = 1}^{n_{\alpha}} \textbf{f} \left[\boldsymbol{\Phi}(\textbf{x}) \textbf{U}_i - \boldsymbol{\Phi}(\textbf{x}') \textbf{U}_j \right] d\textbf{x}' d\textbf{x} = \int_{\Omega(\textbf{x})} \boldsymbol{\Phi}(\textbf{x}) \textbf{f}^{int}(\textbf{x}) d\textbf{x}.
            \end{equation}
            
            In Eq. (\ref{Eq: Matrix Form of Governing Equation in 2D}), $\textbf{M}$ is the mass matrix, and \ref{App: Mass matrix} provides a full derivation of this term.
            In brief, we utilize the lumped mass matrix approach in the present formulation which effectively reduces $\textbf{M}$ to the following expression for each element:
            \begin{equation} \label{Eq: Reduced Mass Matrix}
                \textbf{M} = \frac{m N_{ppe}}{N_{npe}}
            \end{equation}
            where $m$ is the atomic mass, $N_{ppe}$ is the number of particles per element (including lattice points), and $N_{npe}$ is the number of nodes per element \cite{xu2015quasistatic}.
            
            The terms $\Ddot{\textbf{U}}$ and $\textbf{F}^{int}$ are vectors of the respective accelerations and internal forces for each atom/node in the lattice, and $n_{\alpha}$ represents the total number of neighbors of particle $i$ within a specified cutoff radius.
            Furthermore, the force $\textbf{f}^{int}(\textbf{x})$ on particle $i$ at position $\textbf{x}$ is obtained exclusively from the interatomic potential function through relative displacements of particles, and the corresponding net force is obtained through Gaussian integration (see \ref{App: Gaussian integration}). 
            When calculating the force $\textbf{f}^{int}(\textbf{x})$ for a node in the coarse-scaled region, the surrounding lattice points are taken as neighbors.
            The only difference in the fine-scaled force calculations would be the fact that the neighbors of atoms are other atoms rather than lattice points. 
            
    \section{Shock Propagation Technique} \label{Sec: Shock Propagation Technique}

        \subsection{Shock initialization}\label{Sec: Shock initialization}
\begin{figure}[htpb]
                \centering
                \includegraphics[width=0.9\textwidth]{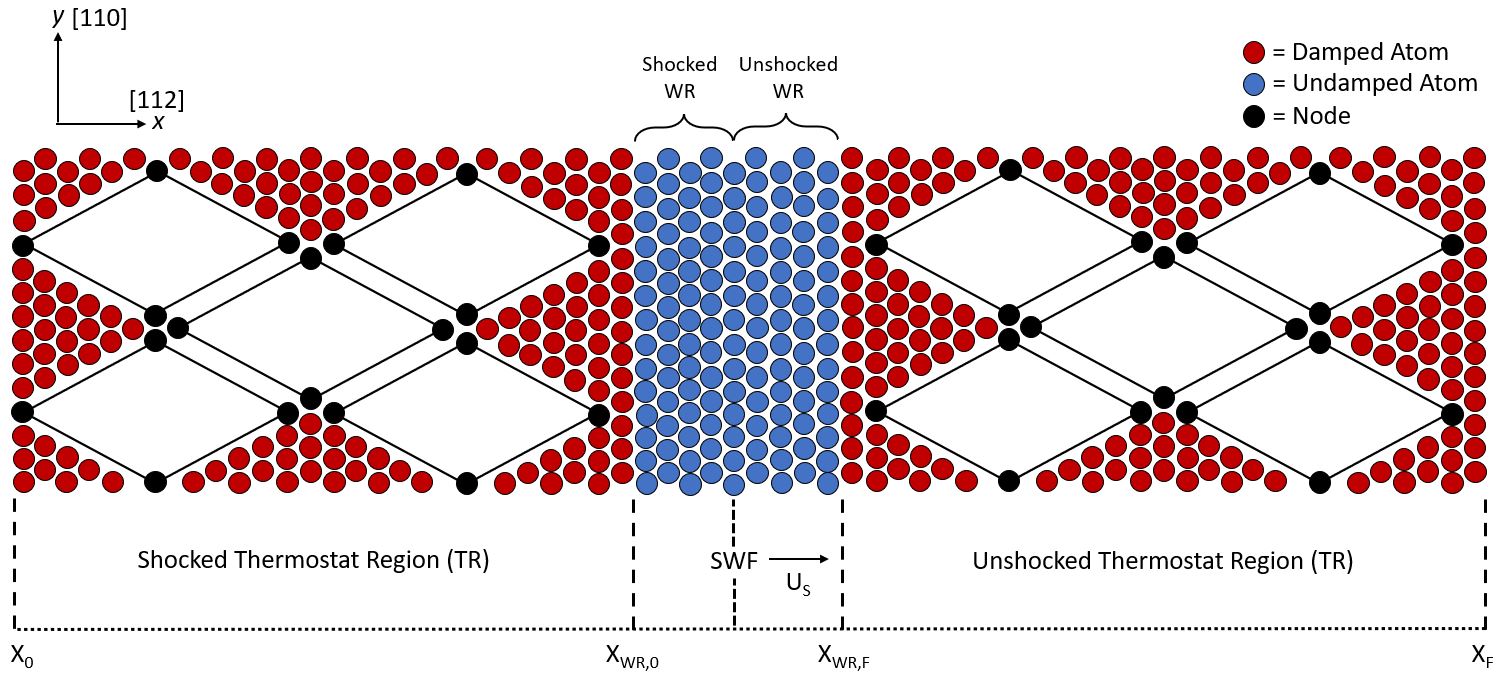}
                \caption{CAC geometry used for shock wave simulations. 
                Here, the red circles represent damped atoms, the blue circles represent undamped atoms, and the black circles represent nodes.}
                \label{Fig:2DCAC_ShockWave_Geometry}
            \end{figure}

            For each simulation, the shock wave is characterized using either the Hugoniot (Sec. \ref{Sec: Hugoniot shock equations}) or Eulerian (Sec. \ref{Sec: Eulerian shock equations}) governing equations, and the shock front is achieved by dividing the grid from Fig. \ref{Fig:2D_CAC_Framework} into different regions as seen in Fig. \ref{Fig:2DCAC_ShockWave_Geometry}.
            The boundary particles within each continuum domain (red circles) constitute the \textit{thermostat regions} (TRs) and are categorized as ``damped" atoms since they apply a constant temperature to the lattice through the Langevin thermostat.
            Furthermore, a small band of inner atoms at each A-C interface are also damped to ensure that the \textit{window region} (WR) made up of ``undamped'' atoms (blue circles) achieves the correct canonical ensemble \cite{davis2022moving}.
            We note that as in \cite{qu2005finite}, the nodes (black circles) are left undamped to prevent spurious behavior within each element.
            The shock wave front (SWF) originates at the center of the WR and travels to the right along the positive $x$-direction with a speed of $U_S$.
            Particles to the right of the SWF constitute the unshocked material while particles to the left constitute the shocked material.
            
            To initialize the shock, we assign a final strain $\epsilon^+$ to the shocked material and use either Eqs. (\ref{Eq: Continuity Equation}) and (\ref{Eq: Linear Law}) for the Hugoniot formulation or Eqs. (\ref{Eq: ClaytonEulerianParticleVelocity}) and (\ref{Eq: ClaytonEulerianShockVelocity}) for the Eulerian formulation to obtain the mean particle velocity $v^+$ and shock front velocity $U_S$.
            The Hugoniot parameters $C_0$ and $S$ as well as the elastic constants $C_{11}$, $\Hat{C}_{111}$, and $\Hat{C}_{1111}$ are initially assigned their literature values given in Table \ref{Table: ShockParameters}.
            The particle velocity $v^+$ represents the new equilibrium velocity for the shocked region, and the strain $\epsilon^+$ causes the lattice to compress uniaxially such that particles behind the SWF obey the Cauchy-Born rule. 
            As a result, the shocked region achieves its final state and the SWF begins to propagate forward starting at the center of the WR. 
            The temperature $\theta^+$ calculated from either Eq. (\ref{Eq: Shock Heat Equation}) or (\ref{Eq: ClaytonEulerianTemperature}) is applied to the shocked TR, and each TR is far enough away from the non-equilibrium SWF to be considered within a region of ``local" equilibrium. 
            Hence, we can legitimately apply the Langevin thermostat to the strained portion of the domain \cite{maillet2000uniaxial}.
            
            In this work, we overcome the runtime-limiting obstacle of boundary reflections present in traditional NEMD shock wave simulations by incorporating two moving window techniques into the multiscale framework.
            The first technique, known as the \textit{conveyor} method, draws inspiration from the moving boundary conditions used in \cite{holland1998ideal} and \cite{selinger2000dynamic} to model dynamic crack propagation as well as the atomic insertion scheme from \cite{Zhakhovskii1997} and \cite{zhakhovsky2011two} to model piston-driven shocks.
            The second technique, known as the \textit{coarsen-refine} method, has similarities to mesh refinement schemes used in finite element \cite{berger1989local,greco2015crack} as well as atomistic-continuum \cite{xu2016mesh,tembhekar2017automatic,amor2021adaptive} frameworks.
            Both techniques serve to track the propagating shock front over engineering length scales and time frames by eliminating shock-boundary reflections, and a description of each can be found in the following sections.
            
        \subsection{Conveyor method} \label{Sec: Conveyor method}
\begin{figure}[htpb]
                \centering
                \includegraphics[width=0.9\textwidth]{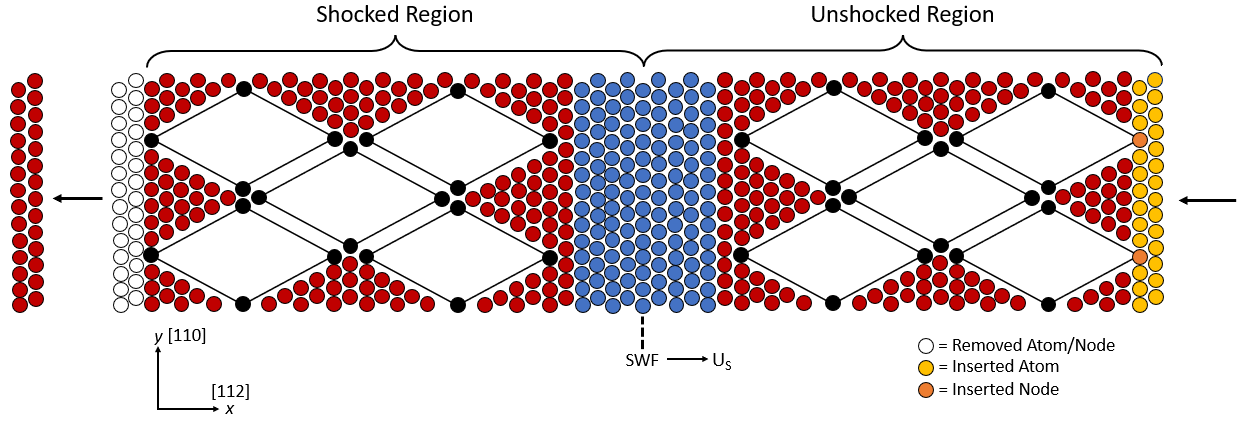}
                \caption{Schematic of the moving window \textit{conveyor} technique for the 2D CAC framework. 
                The white circles represent removed particle locations while the gold/orange circles represent inserted particle locations.}
                \label{Fig:2DCAC_Conveyor_Method}
            \end{figure}

            Figure \ref{Fig:2DCAC_Conveyor_Method} provides a schematic of the conveyor technique for the two-dimensional CAC framework.
            This technique is similar to the scheme found in \cite{davis2022moving} for one dimension, but there are more intricacies and complexities associated with the higher-dimensional lattice. 
            After the SWF has traveled one lattice spacing ($a_{lat}$) along the positive $x$-direction from the center of the WR, the initial position, displacement, velocity, and acceleration of particles in the first two columns of the grid are set equal to the parameters of their rightmost neighbors within the same row.
            The neighbors may be either boundary atoms, nodes, or lattice points, but if they are lattice points, the Verlet parameters are first interpolated as discussed in Sec. \ref{Sec: Finite element implementation}.
            Effectively, the parameters of particles within the first two columns of the lattice are removed from the simulation as is noted in the figure by the leftmost arrow.
            
            This process continues throughout the entire domain from the beginning of the shocked region to the end of the unshocked region, and we note that only the initial position of lattice points are updated since their displacements, velocities, and accelerations are interpolated during the integration algorithm.
            Particles in the final column of the domain (denoted by the gold and orange circles in Fig. \ref{Fig:2DCAC_Conveyor_Method}) are given new initial $x$-positions which are one lattice spacing greater than their current initial $x$-positions, and their $y$-positions remain the same.
            Furthermore, their displacements, velocities, and accelerations are all set equal to zero, and local atomic energy fluctuations induced near $x_F$ are damped by the Langevin thermostat as in \cite{zhakhovsky2011two}.
            This conveyor mechanism occurs with a frequency of $\tau^{-1} = U_{S} / a_{lat}$, and if the simulated and analytical shock velocities are the same, the SWF will remain stationary at the center of the WR for the entire runtime.
            The resulting time resolution of $a_{lat}/U_S$ is thus optimized for the given shock propagation velocity, but higher time resolutions are achievable depending on the speed of the phenomenon in question. 

        \subsection{Coarsen-refine method} \label{Sec: Coarsen-refine method}
\begin{figure}[htpb]
                \centering
                \includegraphics[width=0.9\textwidth]{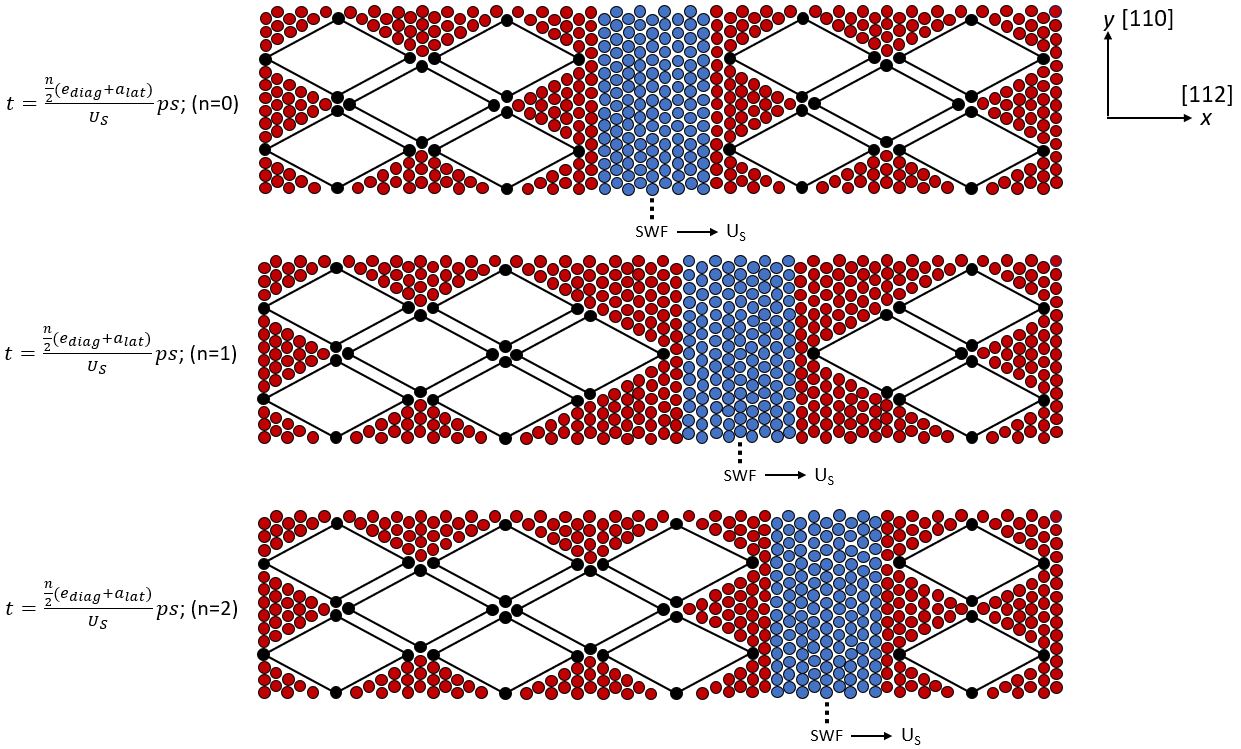}
                \caption{Schematic of the moving window \textit{coarsen-refine} technique for the 2D CAC framework.}
                \label{Fig:2DCAC_CoarsenRefine_Method}
            \end{figure}

            A schematic of the coarsen-refine method can be seen in Fig. \ref{Fig:2DCAC_CoarsenRefine_Method}.
            Here, after the SWF has traveled a distance equal to the length of the element diagonal ($e_{diag}$) plus the lattice spacing divided by two, the moving window mechanism begins whereby material in the shocked continuum region gets coarsened and material in the unshocked continuum region gets refined. 
            In the shocked region, coarsening is achieved by transforming the relevant particles into nodes and lattice points such that new elements appear in the previous atomic locations.
            On the other hand, in the unshocked region, refinement takes place by changing nodes and lattice points into fine-scaled particles through both parameter re-assignment and linear interpolation -- similar to what is done with the conveyor technique.
            This procedure effectively transmits the fine-scaled region forward to the new SWF location as seen in Fig. \ref{Fig:2DCAC_CoarsenRefine_Method}.
            
            After this process completes, undamped particles at the A-C interfaces in the shocked material are redefined as damped particles and vice versa for particles in the unshocked material. 
            Furthermore, the mass matrix is updated to reflect the new mass distribution within the lattice.
            This technique occurs iteratively with a frequency of $\tau^{-1} = U_S / \frac{1}{2} \left(e_{diag}+a_{lat} \right)$, and the integer time counter $n$ is increased by one each time the mechanism terminates (as shown in Fig. \ref{Fig:2DCAC_CoarsenRefine_Method}).
            When utilizing the coarsen-refine method, the entire two-dimensional grid remains stationary and merely the boundaries of the fine-scaled region are modified.
            As a result, most of the domain can be populated with finite elements while a comparatively small section of atoms track the propagating shock wave through the lattice.
            This technique thus ultimately emerges from a consideration of the balance between total efficiency and total accuracy of nonlinear shock wave modeling.
       
    \section{Elastic Anisotropy: Crystal Orientation Dependence on Shock Propagation Response} \label{Sec: Elastic Anisotropy: Crystal Orientation Dependence on Shock Propagation Response}
        In this section, we elaborate on the shock velocity and longitudinal stress results obtained with both the Hugoniot and Eulerian formulations and discuss how they relate to the directional anisotropy of materials subject to shock impact. 
        Recent NEMD works have studied shock propagation along different lattice directions of single crystals and observed a significant orientation dependence on the material's shock response \cite{germann2000orientation,bringa2004atomistic,lin2014effects,neogi2017shock}.
        This phenomenon has also been documented for elastic shock waves in small-scale, atomistic domains \cite{zimmerman2011elastic,davis2020one}.
        Interestingly, large-scale experimental studies have not shown the same orientation dependence of shock parameters \cite{chau2010shock}, but this may be due to the fact that bulk crystals naturally have more defects than what can be feasibly represented using atomistic techniques \cite{lin2014effects}.
        The present work provides a unique insight on this phenomenon because the CAC domain is modeled after the primitive unit cell of an FCC lattice.
        Hence, the shock travels along the [112] longitudinal direction, and the [110] direction is transverse to the direction of propagation.
        To the authors' knowledge, this is one of the first studies to analyze shock evolution along this particular orientation.
        
        \subsection{Simulation specifications} \label{Sec: Simulation specifications}
\begin{figure}[htpb]
                \centering
                \begin{subfigure}{0.9\textwidth}
                    \includegraphics[width=\textwidth]{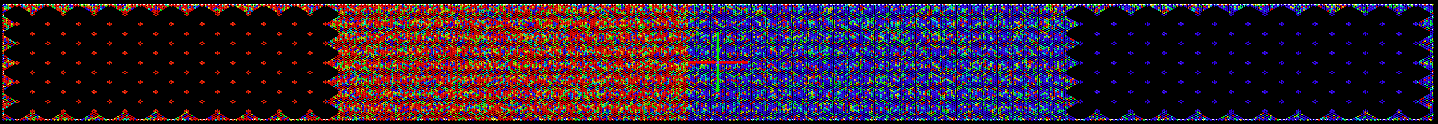}
                    \caption{}
                \end{subfigure}
                \begin{subfigure}{0.45\textwidth}
                    \includegraphics[width=\textwidth]{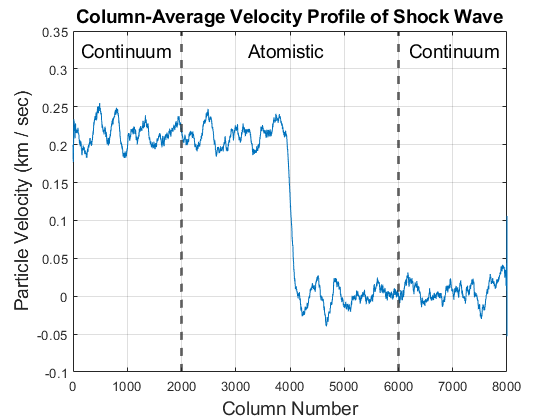}
                    \caption{}
                \end{subfigure}
                \caption{Velocity profiles of the propagating shock in the CAC framework. 
                (a) SWF in the two-dimensional grid (not to scale); (b) SWF obtained from averaging the column velocities of the lattice.}
                \label{Fig:2D_Shock_Conveyor_Simulation}
            \end{figure}

            The results in this section are obtained from shock wave simulations performed with the \textit{conveyor} moving window technique using the CAC domain described in Fig. \ref{Fig:2DCAC_ShockWave_Geometry}. 
            For every simulation, the left and right coarse-scaled regions each contain $250$ particle columns for a total length of $250a_{lat}$, and each element diagonal has a length of $8a_{lat}$.
            Furthermore, the fine-scaled region contains $2500$ particle columns, and the length of each element diagonal is merely the lattice spacing ($a_{lat})$.
            Additionally, each atomistic TR band contains $20$ columns -- much longer than the force range to ensure the WR achieves the desired temperature.
            Simulations are conducted for compressive strains ($\epsilon^+$) ranging from $1\%$ to $9\%$ and $1\%$ to $8\%$ for Cu and Al respectively (see \ref{App: Stress-strain relations}), and the total runtime is $2$ ns.
            A velocity profile of the two-dimensional shocked lattice can be seen in Fig. \ref{Fig:2D_Shock_Conveyor_Simulation}a.
            Specifically, we track the SWF over time in MATLAB by taking a column average of the particle velocities as shown in Fig. \ref{Fig:2D_Shock_Conveyor_Simulation}b.
         
        \subsection{Shock velocity results} \label{Sec: Shock velocity results}
\begin{figure}[htpb]
                \centering
                \begin{subfigure}{0.48\textwidth}
                    \includegraphics[width=\textwidth]{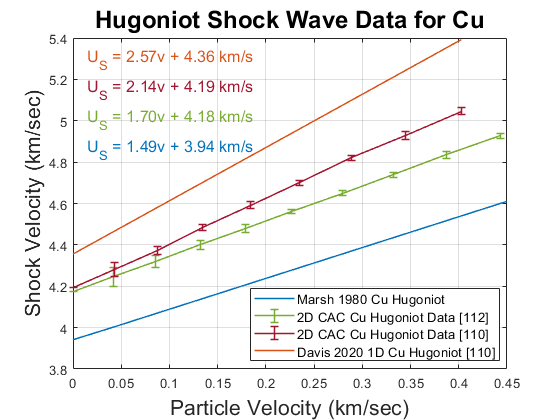}
                    \caption{}
                \end{subfigure}
                \begin{subfigure}{0.48\textwidth}
                    \includegraphics[width=\textwidth]{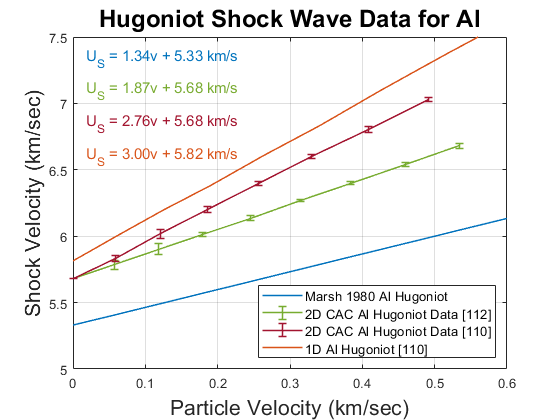}
                    \caption{}
                \end{subfigure}
                \caption{Hugoniot shock wave results for both (a) Cu and (b) Al. 
                The polycrystalline shock Hugoniot obtained from \cite{marsh1980lasl} is shown in blue. 
                Two-dimensional CAC Hugoniot data obtained for shocks propagating along the [112] and [110] lattice directions are shown in green and red respectively. 
                One-dimensional shock Hugoniots are given in orange. 
                The Cu Hugoniot comes from \cite{davis2020one}, and the Al Hugoniot is calculated in the present work.}
                \label{Fig:Shock_Hugoniot_Conveyor_Velocity}
            \end{figure}

\begin{figure}[htpb]
                \centering
                \begin{subfigure}{0.48\textwidth}
                    \includegraphics[width=\textwidth]{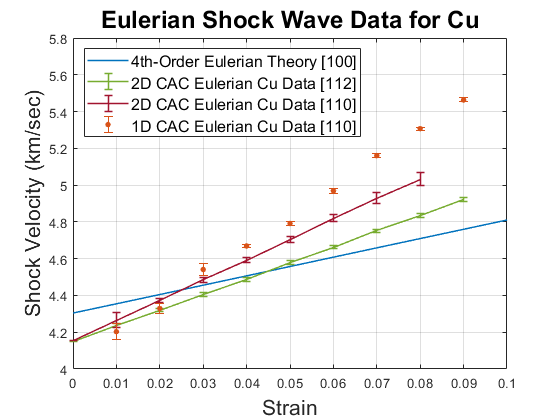}
                    \caption{}
                \end{subfigure}
                \begin{subfigure}{0.48\textwidth}
                    \includegraphics[width=\textwidth]{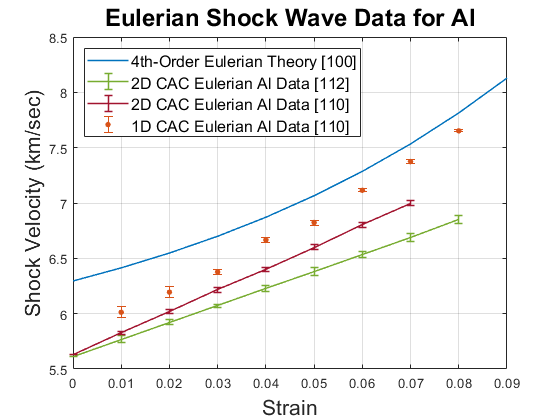}
                    \caption{}
                \end{subfigure}
                \caption{Eulerian shock results for both (a) Cu and (b) Al. 
                The blue line represents velocities obtained from fourth-order Eulerian theory. 
                Two-dimensional CAC data obtained for shocks propagating along the [112] and [110] lattice directions are shown in green and red respectively.
                One-dimensional CAC data obtained from \cite{davis2022moving} are in orange.}
                \label{Fig:Shock_Eulerian_Conveyor_Velocity}
            \end{figure}

            Shock velocity results obtained for both Hugoniot and Eulerian theory can be seen in Figs. \ref{Fig:Shock_Hugoniot_Conveyor_Velocity} and \ref{Fig:Shock_Eulerian_Conveyor_Velocity} respectively.
            Specifically, Fig. \ref{Fig:Shock_Hugoniot_Conveyor_Velocity} displays the shock velocity vs. particle velocity data (as well as the derived Hugoniot equations) of four different sets of simulations using both (a) Cu and (b) Al.
            Here, the blue line represents the polycrystalline Hugoniot calculated in \cite{marsh1980lasl}, and the green data points are the average velocity results for shocks propagating through the standard CAC domain.
            As a comparison, we also invert the lattice such that the [110] orientation lies along the $x$-direction, and the [112] orientation lies along the $y$-direction, and these results are given by the red data points.
            As in \ref{App: Stress-strain relations}, we performed stress vs. strain studies for this inverted lattice and found yielding to occur at 9\% strain for Cu and 8\% strain for Al, so we maintain $\epsilon^+$ values below these elastic limits when simulating shocks along the [110] direction.
            Finally, we also present one-dimensional atomistic shock data obtained from \cite{davis2020one} for Cu and calculated in this work for Al. 
            
            The data and associated Hugoniot equations in Fig. \ref{Fig:Shock_Hugoniot_Conveyor_Velocity} clearly show the dependency of a shock's propagation velocity on the given lattice orientation.
            In particular, both of the two-dimensional CAC Hugoniots have $C_0$ and $S$ values which are greater than the standard polycrystalline Hugoniot.
            This is most likely due to the fact that the FCC primitive unit cell is rhombohedral instead of cubic, so the entire CAC lattice is more compressed than a traditional structured FCC grid.
            This causes the particles in the domain to be more compact which results in larger forces from the interatomic potential and hence higher shock velocities.
            As expected, the inverted CAC lattice produces slightly higher shock velocities than the lattice from Fig. \ref{Fig:2DCAC_ShockWave_Geometry} since the [110] lattice spacing is shorter than the [112] spacing.
            Finally, the one-dimensional shock velocities are greater than the those from the two-dimensional simulations due to the lack of any transverse motion which naturally dampens the shock speed. 
            Instead, the 1D results are comparable to plane-plane collisions in a bulk lattice \cite{tsai1966shock}.
            
            We observe a similar phenomenon for the Eulerian results in Fig. \ref{Fig:Shock_Eulerian_Conveyor_Velocity} where we now plot average shock velocity vs. applied strain.
            Here, the green and red data points are from the same types of 2D simulations as those from Fig. \ref{Fig:Shock_Hugoniot_Conveyor_Velocity}. 
            However, the blue line now represents the analytical results from fourth-order Eulerian theory, and the orange data points are 1D CAC shock results obtained from \cite{davis2022moving}. 
            As seen previously, the 1D shock velocities are slightly greater than the 2D velocities from the present study, and the inverted CAC lattice has a higher slope than the standard CAC lattice. 
            For Cu, the shock velocities predicted at higher strains by Eulerian theory are indeed lower than the 2D and 1D CAC results, but this is not the case for Al. 
            The reason for the anomalous results with Al is not necessarily clear, but it could be due to the higher presence of aluminum alloys which could alter the density and thus affect the model's outcomes.
            Nonetheless, we observe qualitative compatibility between the Hugoniot and Eulerian formulations which gives us confidence that the current CAC framework produces accurate results and can thus be reliably used to measure the response of materials to shock propagation along various lattice directions.
            
        \subsection{Longitudinal stress results} \label{Sec: Longitudinal stress results}
\begin{figure}[htpb]
                \centering
                \begin{subfigure}{0.48\textwidth}
                    \includegraphics[width=\textwidth]{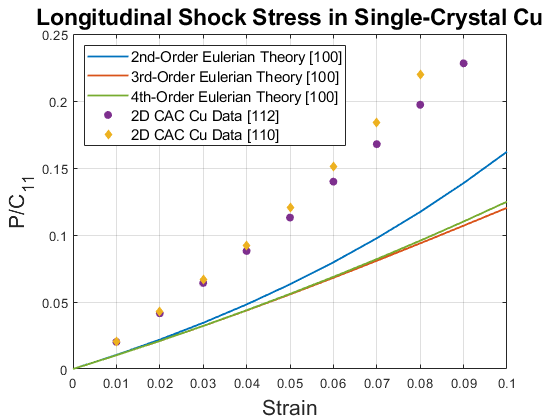}
                    \caption{}
                \end{subfigure}
                \begin{subfigure}{0.48\textwidth}
                    \includegraphics[width=\textwidth]{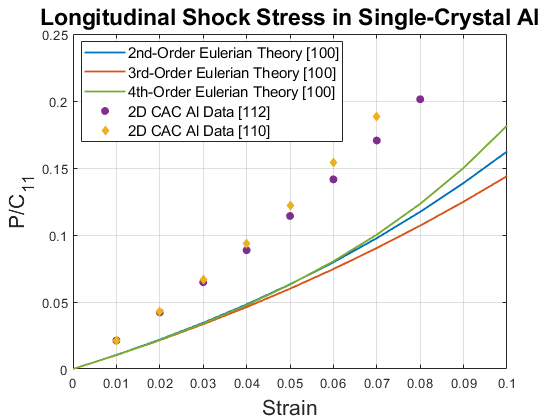}
                    \caption{}
                \end{subfigure}
                \caption{Longitudinal stress data for both (a) Cu and (b) Al. 
                The blue, orange, and green lines represent the [100] results from 2nd, 3rd, and 4th-order Eulerian theory respectively.
                The purple circles and gold diamonds represent the [112] and [110] CAC data respectively.}
                \label{Fig:Longitudinal_Shock_Stress}
            \end{figure}

            To supplement the anisotropic shock velocity results from Sec. \ref{Sec: Shock velocity results}, we perform longitudinal stress vs. strain studies using the shocked data for both Cu and Al, and these results can be seen in Fig. \ref{Fig:Longitudinal_Shock_Stress}.
            Specifically, we calculate the time-averaged virial (thermodynamic) stress ($\sigma_{xx}$) in the shocked region using Eq. (\ref{Eq: Virial Stress}), and we relate the Cauchy stress ($P_{xx}$) to the virial stress as follows \cite{zimmerman2011elastic}:
            \begin{equation} \label{Eq: Cauchy Stress}
                P_{xx} = (1-\epsilon)\sigma_{xx}
            \end{equation}
            where we note that compressive stress/strain is considered positive.
            
            Fig. \ref{Fig:Longitudinal_Shock_Stress} shows the shock stress $P_{xx}$ normalized by the second-order elastic constant $C_{11}$ as a function of the applied strain.
            The data from Hugoniot and Eulerian theory were practically identical, so without loss of generality, we only exhibit the Eulerian results.
            The [100] second, third, and fourth-order Eulerian models are represented by the blue, orange, and green lines respectively, while the [112] and [110] shock stress data are represented by the purple circles and gold diamonds respectively.
            As in Sec. \ref{Sec: Shock velocity results}, we clearly observe the orientation dependence of the shock stress as the CAC data is significantly higher than that predicted by the various Eulerian models for shocks along the [100] direction.
            Furthermore, the [110] CAC simulations produced shock stresses which were slightly higher than those from the [112] simulations.
            Again, this is primarily due to the higher compression velocities caused by the larger `compactness' of CAC domains.
            This anisotropic stress data is congruent with a previous work which analyzed elastic shocks along various lattice directions using a number of different potential functions \cite{zimmerman2011elastic}.
    
    \section{Results with the Coarsen-Refine Method and Formulation Efficiency} \label{Sec: Results with the Coarsen-Refine Method and Formulation Efficiency}
            Without loss of generality, we only reference data from Eulerian theory in this section as both shock models gave similar quantitative results.
        
        \subsection{Coarsen-refine simulations}
\begin{figure}[htpb]
                \centering
                \includegraphics[width=0.9\textwidth]{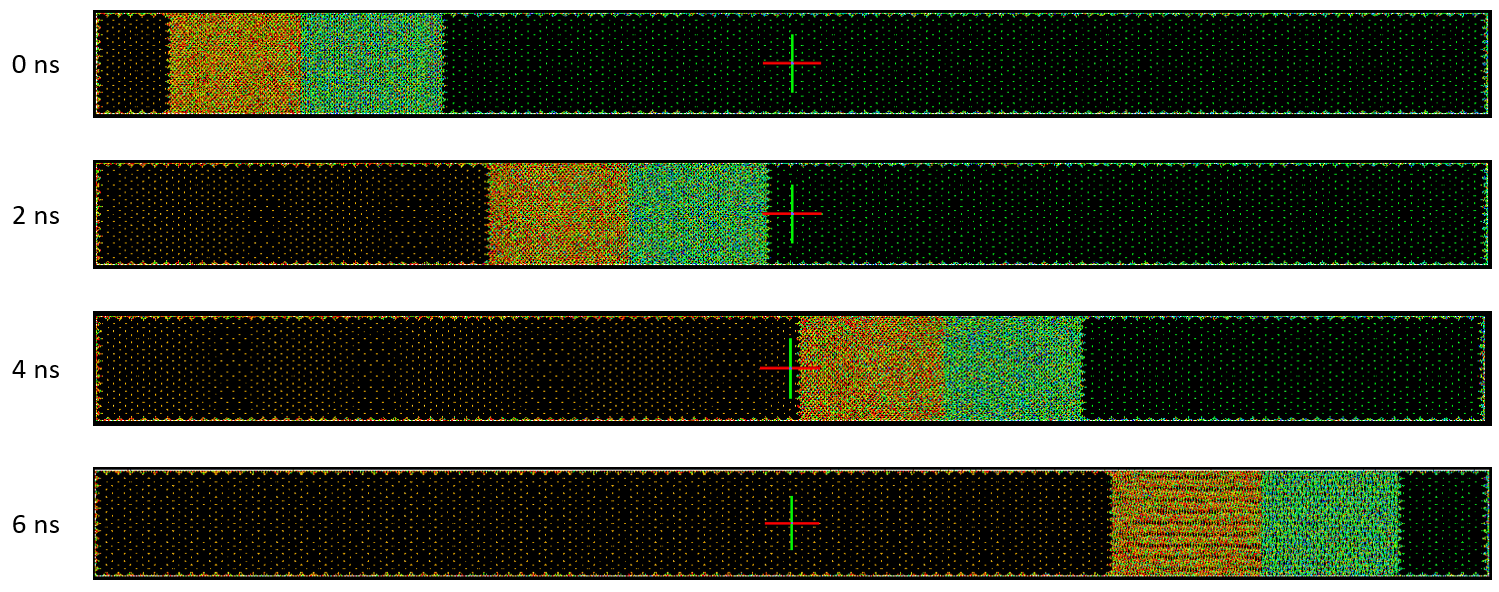}
                \caption{Shock simulation using the coarsen-refine moving window technique.}
                \label{Fig:Coarsen_Refine_2D_Simulation}
            \end{figure}

            In Fig. \ref{Fig:Coarsen_Refine_2D_Simulation}, we present results from a shock wave simulation performed using the \textit{coarsen-refine} technique over $6$ ns. 
            Here, we can observe the atomistic portion of the domain successfully follow the evolving shock front throughout the CAC framework with no spurious wave behavior at the A-C interfaces. 
            Due the elastic nature of the shock as discussed in Sec. \ref{Sec: Elastic shock waves}, no dislocations are present to the left of the wave front, but we do see the shocked material maintain the mean particle velocity of $v^+$ for the entire runtime.
            These results are in contrast to those performed using the conveyor technique because now the SWF may travel through the entire CAC domain while staying within the fine-scaled region.
            Although previous work has used mesh refinement to study phenomena within both finite-element \cite{berger1989local} and multiscale \cite{xu2016mesh,tembhekar2017automatic,amor2021adaptive} schemes, utilizing simultaneous refine/coarsen techniques to study dynamic, high-temperature phenomena is still a challenging area of research \cite{davis2022moving}. 
            Thus, the present formulation provides a novel means for tracking propagating shocks over long runtimes, and may be used to research even more complex lattice structures in the future such as nanoscale composites or high-entropy alloys.
        
        \subsection{Shock structure and planarity}
\begin{figure}[htpb]
                \centering
                \includegraphics[width=0.5\textwidth]{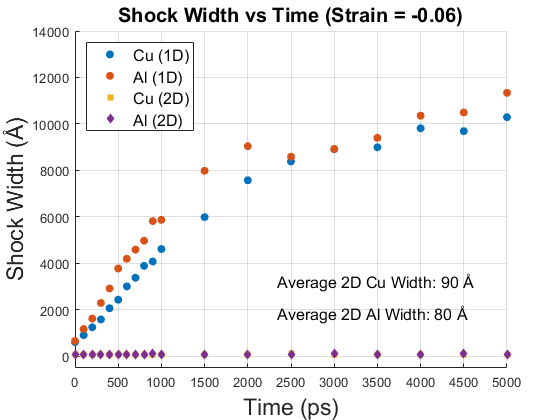}
                \caption{Spatial shock width over time. 
                The blue and red circles represent the 1D CAC data from \cite{davis2022moving} for both Cu and Al respectively.
                The gold squares and purple diamonds represent the 2D CAC data for Cu and Al from the present work.}
                \label{Fig:2D_Shock_Width}
            \end{figure}

            We now use the coarsen-refine simulations to analyze the shock front's spatial width over $5$ ns, and the results for $\epsilon^+ = -0.06$ can be seen in Fig. \ref{Fig:2D_Shock_Width}.
            As a comparison, we also show the 1D CAC results from \cite{davis2022moving}.
            Unlike the 1D data, the present work shows a clear steadiness in the shock wave behavior as evidenced by the fact that the shock width remains constant throughout the simulation with very little deviation from the mean.
            We also do not observe a significant change in the shock front's planarity throughout the simulation's duration.
            Finally, similar results were found for both Cu and Al over the range of strains studied with the present formulation.
            Clearly, for shock waves modeled at the microscale, the ability of particles to oscillate transversely to the direction of shock propagation plays a large role in the overall steadiness of the wave.
            These results are similar to findings from previous NEMD studies which observed a change in shock structure and steadiness when transitioning from a 1D to 3D regime \cite{holian1995atomistic}.
            In particular, the transition from unsteady to steady waves was due to the ``increase in coupling between vibrational excitations normal and transverse to the direction of shock wave propagation'' \cite{holian1979molecular}.
            Our work shows this for two dimensions as well.
        
        \subsection{Framework speedup and efficiency}
\begin{figure}[htpb]
                \centering
                \begin{subfigure}{0.48\textwidth}
                    \includegraphics[width=\textwidth]{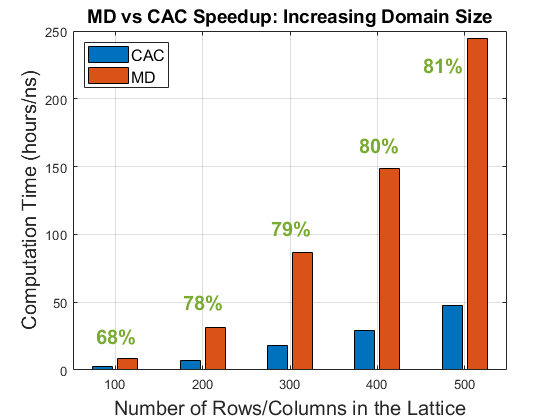}
                    \caption{}
                \end{subfigure}
                \begin{subfigure}{0.48\textwidth}
                    \includegraphics[width=\textwidth]{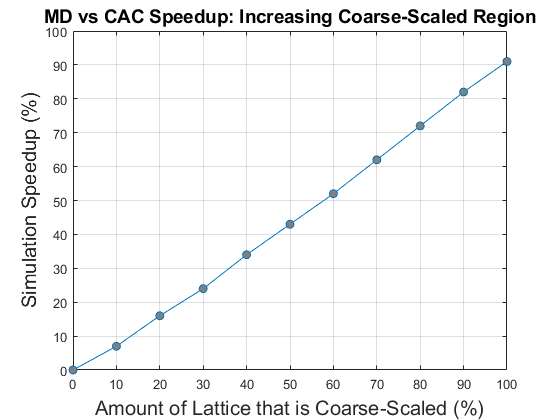}
                    \caption{}
                \end{subfigure}
                \caption{Efficiency of the CAC framework vs equally-sized MD domains.  
                In (a), the total runtimes are compared for increasing system sizes. 
                Here, the central fine-scaled region of the CAC lattice is always 1/10 the length of the entire grid.
                In (b), the simulation speedup is shown when the size of the domain remains constant, but the coarse-scaled region increases from 0\% to 100\% of the lattice.}
                \label{Fig:Domain_Size_Speedup}
            \end{figure}

            For the sake of completeness, we now present results for speedup/efficiency tests which compare the two-dimensional moving window CAC framework to equally-sized NEMD domains.
            The data from these two studies can be seen in Fig. \ref{Fig:Domain_Size_Speedup}.
            Specifically, in Fig. \ref{Fig:Domain_Size_Speedup}a, we maintain a constant ratio in the CAC lattice such that the fine-scaled region is always one-tenth the length of the entire grid, and we run simulations for increasing domain sizes. 
            We observe the CAC vs. MD efficiency reach an asymptotic value around 81\% (further increases in domain size did not significantly effect the speedup percentage).
            Next, in Fig. \ref{Fig:Domain_Size_Speedup}b, we keep the total lattice size constant and vary the length of the coarse-scaled region from 0\% to 100\% of the total area. 
            Clearly, as the percentage of the lattice that is coarse-scaled increases, the speedup does as well, and we note that this increase appears to be fairly linear. 
            These studies demonstrate the utility of using the present CAC framework to enhance performance in large-scale simulations.

    \section{Conclusion} \label{Sec: Conclusion}
        In this paper, we developed a dynamic moving window CAC framework to simulate shock wave propagation through a two-dimensional, single-crystal lattice. 
        Specifically, we characterized the shock using both the linear Hugoniot \cite{davison2008fundamentals} and nonlinear Eulerian \cite{clayton2013nonlinear} shock equations to study the classic Riemann problem of a single discontinuity traveling through an infinite medium.
        The CAC multiscale formulation was utilized for its ability to seamlessly transition between the fine-scaled and coarse-scaled regions, and many verifications and analyses were conducted on the higher-dimensional system.
        We elaborated on the technique to initialize the shock front in the lattice as well as described two moving window methods which were incorporated into the domain.
        These schemes provided a mechanism to study the evolution of the shock over very long simulation times by preventing non-physical wave reflections at the A-C interfaces.
        
        We performed many shock wave simulations within the CAC framework and used the moving window techniques to track the shock front through two different FCC materials: Cu and Al. 
        The unique lattice directions inherent to the CAC formulation provided us the opportunity to study how directional anisotropies in single crystals can give rise to orientation-dependent shock velocities.
        We observed that longitudinal shocks traveling along the [112] and [110] directions of the CAC domain propagated at distinct velocities for a given strain and particle velocity. 
        These shock velocities were also different from those predicted by polycrystalline Hugoniot and Eulerian analytical models as well as previous one-dimensional atomistic and multiscale data. 
        From these results, we were able to derive new Hugoniot parameters for the CAC formulation, and longitudinal stress calculations further validated the observed anisotropic material response.
        Our data agreed qualitatively with the results from previous NEMD studies which identified this orientation-dependence of shock evolution in solids \cite{germann2000orientation,bringa2004atomistic,lin2014effects,neogi2017shock}.
        
        Next, in Sec. \ref{Sec: Results with the Coarsen-Refine Method and Formulation Efficiency}, we exhibited the capability and novelty of the present framework by using the coarsen-refine technique to track a propagating shock wave through the entire grid.
        By leveraging concepts from previous atomistic and finite element schemes as well as exploiting the unique qualities of the CAC formulation, the fine-scaled region could travel through the domain at the speed of the moving wave front, and we noted the significance of this for advancing non-equilibrium multiscale research.
        We utilized this techinique to study the shock's structure and planarity over very long runtimes which are typically unattainable in traditional NEMD methods.
        Finally, we presented multiple plots comparing the efficiency of an NEMD system to an equally-sized CAC lattice.
        We observed that the present moving window multiscale scheme had significantly faster runtimes for various domain sizes -- a necessary quality for realistic and scalable atomistic-continuum models.
        
        The present work is innovative in its own right, but it also opens the door to more complex research involving the use of multiscale domains to simulate dynamic, nonlinear phenomena over engineering length scales.
        While we focused only on elastic shock waves in this work, we hope to expand this formulation to model elastic-plastic shocks \cite{lloyd2014simulation} in polycrystalline materials to study the role of grain boundaries on shock evolution.
        Additionally, recent works have used both atomistic \cite{shen2022uncovering,jiang2022molecular} as well as multiscale \cite{chu2022multiscale,elahi2022multiscale} methods to predict material behavior in medium-entropy and high-entropy alloys.
        This work provides a framework to study shock propagation through such materials. 
        Furthermore, we would also like to utilize machine learning algorithms in this scheme to pass information from the mesoscale to macroscale \cite{xiao2021machine}.
        Finally, we hope to incorporate a high-frequency wave passing technique that was first introduced in \cite{chen2018passing} and \cite{DAVIS2022111702} into the present formulation to study shock scattering and the role of scattered waves in subsequent material behavior.
        
    \section{Acknowledgments} \label{Sec: Acknowledgments}
        This material is based upon work supported by the National Science Foundation under Grant No. $1950488$.
        Financial support was also provided by the U.S. Department of Defense through the National Defense Science and Engineering Graduate (NDSEG) Fellowship Program (F-$1656215698$). 
        Simulations were performed using the Easley computing cluster at Auburn University.

    \bibliographystyle{ieeetr}
    \bibliography{CAC2D_MW_Shocks}
    
    \appendix
    
    \section{Verifications} \label{Sec: Verifications}
        In this section, we present results from additional studies which verify that the current CAC framework functions correctly.
    
        \subsection{Temperature equilibration} \label{App: Temperature equilibration}
\begin{figure}[htpb]
                \centering
                \begin{subfigure}{0.48\textwidth}
                    \includegraphics[width=\textwidth]{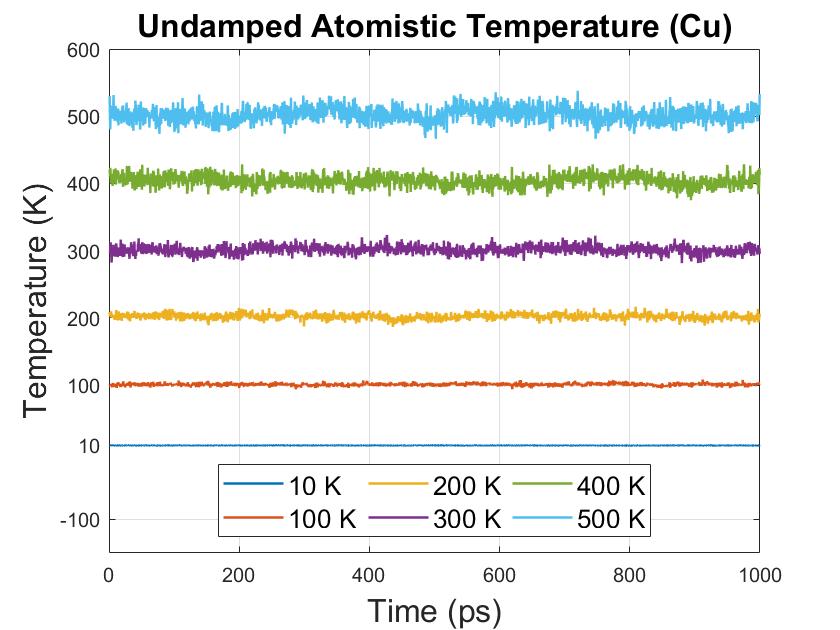}
                    \caption{}
                \end{subfigure}
                \begin{subfigure}{0.48\textwidth}
                    \includegraphics[width=\textwidth]{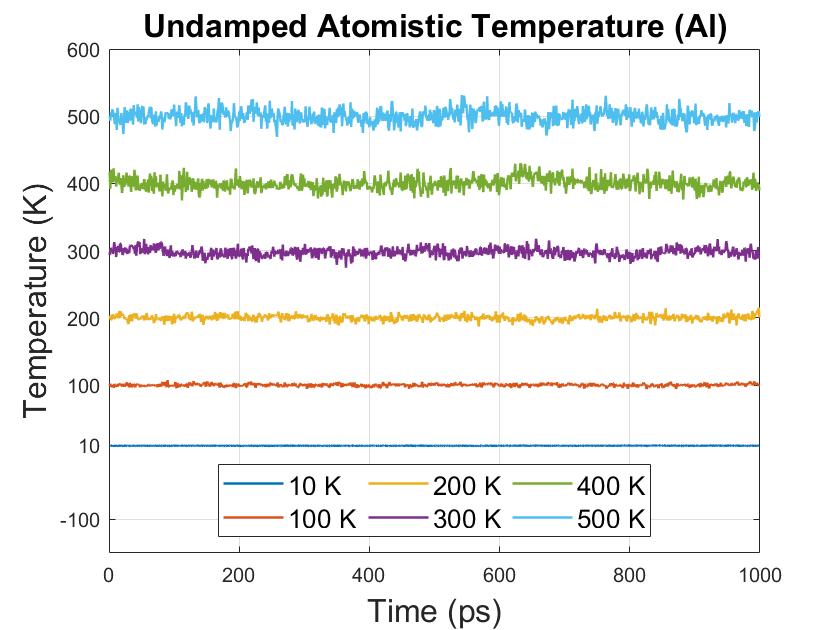}
                    \caption{}
                \end{subfigure}
                \caption{Temperature in the undamped atomistic region of the CAC framework vs. time using the $Morse$ potential for both (a) Cu and (b) Al.
                The $Langevin$ thermostat is applied to the TRs for the following input temperatures: 10 K, 100 K, 200 K, 300 K, 400 K, and 500 K.}
                \label{Fig:UndampedAtomisticTemperature}
            \end{figure}

            First, we verify that the two-dimensional CAC framework used in the shock wave simulations (Fig. \ref{Fig:2DCAC_ShockWave_Geometry}) can achieve the correct canonical ensemble in the undamped WR when the Langevin thermostat is applied to each TR. 
            In particular, we demonstrate that the system equilibrates to the proper steady-state value over long simulation times for a range of input temperatures.
            The initial random velocities of the particles are such that the system has the correct total energy for a given temperature $\theta_0$.
            Furthermore, we ensure that each atomistic TR has a length which is at least equal to the force range of the interatomic potential, and we set the damping parameter $\zeta$ equal to one-half the Debye frequency of the material ($\frac{1}{2} \omega_D$).
            These specifications are based off results from previous multiscale studies which used CADD \cite{qu2005finite} as well as CAC \cite{davis2022moving} to characterize the domain.
            The temperature equilibration results for both Cu and Al can be seen in Fig. \ref{Fig:UndampedAtomisticTemperature}.
            
            The domain size for these simulations is as follows: $200$ total columns ($100$ in the atomistic region and $50$ in each continuum region) and $40$ total rows. 
            Within the fine-scaled region, there are $5$ columns in each TR and hence $90$ columns in the undamped WR. 
            Each simulation is performed for $1$ ns, and the temperature in the WR is obtained at every time step using the equipartition theorem.
            As stated in Sec. \ref{Sec: Integration algorithm and thermostat}, the maximum temperature obtained from shock loading in this work is $\sim$ 450 K, so we perform analysis for the following temperatures: $10$ K, $100$ K, $200$ K, $300$ K, $400$ K, and $500$ K.
            In Fig. \ref{Fig:UndampedAtomisticTemperature}, we observe that in each simulation, the temperature achieves a steady state around its mean value with very little deviation.
            Hence, this confirms the implementation of the framework from Fig. \ref{Fig:2DCAC_ShockWave_Geometry} and shows that the WR can maintain the correct equilibrium temperature during long runtimes with both materials.
            
        \subsection{Stress-strain relations} \label{App: Stress-strain relations}
\begin{figure}[htpb]
                \centering
                \begin{subfigure}{0.48\textwidth}
                    \includegraphics[width=\textwidth]{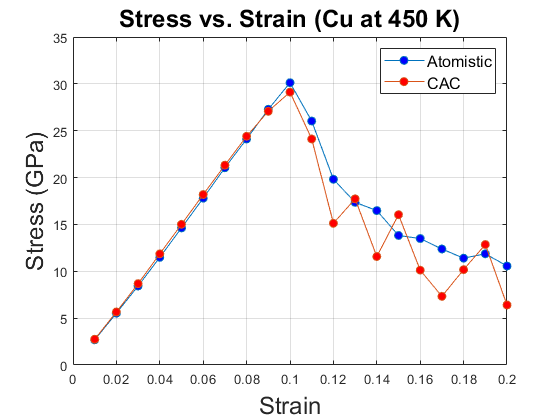}
                    \caption{}
                \end{subfigure}
                \begin{subfigure}{0.48\textwidth}
                    \includegraphics[width=\textwidth]{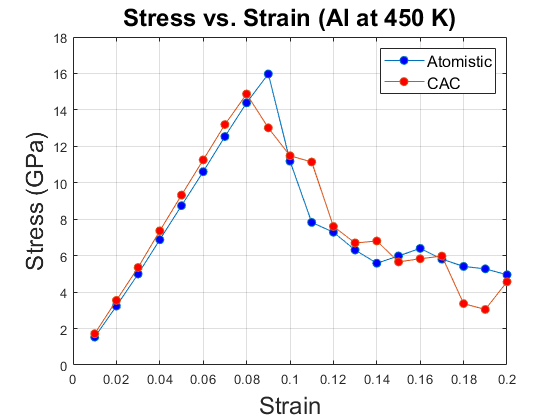}
                    \caption{}
                \end{subfigure}
                \caption{Virial stress of the domain as a function of strain for both (a) Cu and (b) Al.
                Simulations were performed for both a purely atomistic (blue) and CAC (red) framework.
                In each case, the system is equilibrated to $500$ K, and the compression is applied uniaxially along the $x$-direction.}
                \label{Fig:StressVsStrain}
            \end{figure}

            Next, we identify the elastic zone of the framework and ensure that the yield strength between a purely atomistic domain and equally-sized CAC domain is comparable.
            This is done to establish that the CAC force calculations are accurate as well as provide a range of input strains for the shock equations.
            Specifically, we compress the grid uniaxially along the $x$-direction ([112] lattice orientation) with strains ranging from $0.01 - 0.2$ and calculate the virial stress of the domain for each input strain using the following expression \cite{tadmor2011modeling}:
            \begin{equation} \label{Eq: Virial Stress}
                \sigma_{kl} = \frac{1}{A} \left< -\sum_{\alpha} m^{\alpha} \left(\Dot{u}_k^{\alpha} - \Bar{\Dot{u}}_k\right) \left(\Dot{u}_l^{\alpha} - \Bar{\Dot{u}}_l\right)  + \frac{1}{2} \sum_{\substack{\alpha, \beta{}\\(\alpha \ne \beta)}} \varphi^{\alpha \beta} \frac{r_k^{\alpha \beta} r_l^{\alpha \beta}}{r^{\alpha \beta}} \right>.
            \end{equation}
            In Eq. (\ref{Eq: Virial Stress}), $\boldsymbol{\sigma}$ is the virial (thermodynamic) stress, $A$ is the area of the grid, $m^{\alpha}$ is the mass of particle $\alpha$, $\Dot{u}_k^{\alpha}$ is the velocity in the $k^{th}$ direction of particle $\alpha$, $\Bar{\Dot{u}}_k$ is the average velocity in the $k^{th}$ direction of all particles in the given area, $\varphi^{\alpha \beta}$ is the first derivative of the potential energy at a distance $r^{\alpha \beta}$ between particles $\alpha$ and $\beta$ ($\varphi^{\alpha \beta} = \frac{\partial \Pi}{\partial r^{\alpha \beta}}$), and $r_k^{\alpha \beta}$ is the distance in the $k^{th}$ direction between particles $\alpha$ and $\beta$.
            Since we consider uniaxial compressive strains for the shock simulations, we only calculate the longitudinal stress ($\sigma_{kk} = \sigma_{xx}$) in this section and do not perform any tensile tests.
            The stress vs. strain results for both Cu and Al at $450$ K can be seen in Fig \ref{Fig:StressVsStrain}.
            
            For each atomistic simulation, the domain contains $100$ columns and $20$ rows, and the runtime is $100$ ps with an equilibration time of $50$ ps. 
            The parameters for the CAC simulations are the same, but the lattice described in Sec. \ref{Sec: Geometry and boundary conditions} (with damped atoms) is utilized instead of the fully atomistic grid.
            Since $450$ K is the highest temperature achieved in the shock simulations, we specifically wanted to identify the yield point at this extreme temperature to inform our shock calculations.
            In Fig. \ref{Fig:StressVsStrain}, we observe that the linear elastic region of the CAC framework is nearly identical to that of the atomistic framework for both Cu and Al with yielding occurring at compressive strains of approximately $10\%$ and $9\%$ respectively.
            After this point, dislocations begin appearing throughout the lattice in both the coarse-scaled and fine-scaled regions, so we maintain compressive strains $\le$ $9\%$ for Cu and $\le$ $8\%$ for Al when using the Hugoniot and Eulerian shock equations.
            These results confirm the validity of the CAC force calculations, and they are also congruent with results from previous CAC studies \cite{xiong2011coarse,xiong2012concurrent,xiong2012coarse}.

    \section{Additional Information on 2D CAC Elements} \label{App: Additional Information on 2D CAC Elements}
        
        \subsection{Mass matrix} \label{App: Mass matrix}
            We now elaborate on the isoparametric formulation of the mass matrix for a given continuum element in the 2D CAC framework.
            As stated in  Sec. \ref{Sec: Two-dimensional formulation}, element connectivity is not required in CAC. 
            Hence, this derivation is general and can be applied to any element in the domain assuming the physical nodal coordinates of that element are known.
            
\begin{figure}[htpb]
                \centering
                \includegraphics[width=0.75\textwidth]{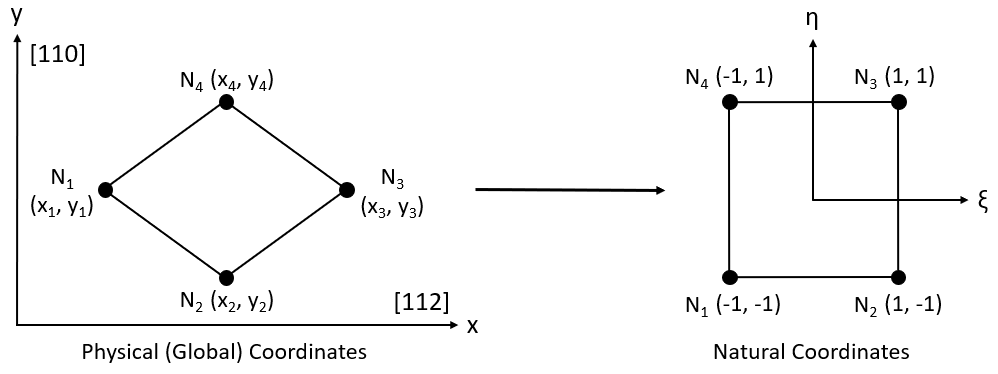}
                \caption{Mapping from global to natural coordinates of a two-dimensional CAC element.}
                \label{Fig:GlobalToNaturalCoordinates}
            \end{figure}

            The isoparametric shape functions of a four-node element are given as follows:
            \begin{align}
                \phi_1(\xi, \eta) &= \frac{1}{4}(1-\xi)(1-\eta) \\
                \phi_2(\xi, \eta) &= \frac{1}{4}(1+\xi)(1-\eta) \\
                \phi_3(\xi, \eta) &= \frac{1}{4}(1-\xi)(1+\eta) \\
                \phi_4(\xi, \eta) &= \frac{1}{4}(1+\xi)(1+\eta)
            \end{align}
            which can be stored in a matrix as
            \begin{equation}
                \boldsymbol{\Phi}(\xi, \eta) = 
                \begin{bmatrix}
                    \phi_1 & 0 & \phi_2 & 0 & \phi_3 & 0 & \phi_4 & 0 \\
                    0 & \phi_1 & 0 & \phi_2 & 0 & \phi_3 & 0 & \phi_4 
                \end{bmatrix}
            \end{equation}
            where $\phi_i = \phi_i(\xi, \eta)$.
            In order to map the element between the global and natural coordinate system, we need the Jacobian which is given as follows:
            \begin{equation} \label{Eq: Jacobian}
                \boldsymbol{J} = 
                \begin{bmatrix}
                    \partial \phi_1 / \partial \xi & \partial \phi_2 / \partial \xi & \partial \phi_3 / \partial \xi & \partial \phi_4 / \partial \xi \\
                    \partial \phi_1 / \partial \eta & \partial \phi_2 / \partial \eta & \partial \phi_3 / \partial \eta & \partial \phi_4 / \partial \eta
                \end{bmatrix}
                \begin{bmatrix}
                    x_1 & y_1 \\
                    x_2 & y_2 \\
                    x_3 & y_3 \\
                    x_4 & y_4
                \end{bmatrix}
            \end{equation}
            where ($x_1, y_1$), ($x_2, y_2$), ($x_3, y_3$), and ($x_4, y_4$) are the positions of the four element nodes in the global coordinate system.
            We note that the numbering goes counterclockwise starting from the left node as seen in Fig. \ref{Fig:GlobalToNaturalCoordinates}.
            Expanding out Eq. (\ref{Eq: Jacobian}) and taking the appropriate derivatives of the shape functions, we obtain the four components of the Jacobian:
            \begin{align}
                J_1 &= \frac{x_1}{4}(\eta-1) + \frac{x_2}{4}(1-\eta) + \frac{x_3}{4}(\eta+1) - \frac{x_4}{4}(\eta+1) \\
                J_2 &= \frac{y_1}{4}(\eta-1) + \frac{y_2}{4}(1-\eta) + \frac{y_3}{4}(\eta+1) - \frac{y_4}{4}(\eta+1) \\
                J_3 &= \frac{x_1}{4}(\xi-1) - \frac{x_2}{4}(\xi+1) + \frac{x_3}{4}(\xi+1) + \frac{x_4}{4}(1-\xi) \\
                J_4 &= \frac{y_1}{4}(\xi-1) - \frac{y_2}{4}(\xi+1) + \frac{y_3}{4}(\xi+1) + \frac{y_4}{4}(1-\xi).
            \end{align}
            Hence, the Jacobian determinant is
            \begin{equation} \label{Eq: detJ}
                det (\textbf{J}) = 
                \begin{vmatrix}
                    J_1 & J_2 \\
                    J_3 & J_4
                \end{vmatrix} = 
                J_1 J_4 - J_2 J_3
            \end{equation}
            which can be simplified using a software program like Wolfram Mathematica.
            
            The expression for the mass matrix of the 2D element in global coordinates is given as follows:
            \begin{equation}
                \textbf{M} = \rho \int_A \boldsymbol{\Phi}^T \boldsymbol{\Phi} \,\, dA.
            \end{equation}
            Writing this in natural coordinates:
            \begin{equation}
                \textbf{M} = \rho \int_{-1}^{1} \int_{-1}^{1} \left[ \boldsymbol{\Phi}^T \boldsymbol{\Phi} \cdot det (\textbf{J}) \right] d\xi d\eta
            \end{equation}
            where 
            \begin{equation}
                \boldsymbol{\Phi}^T \boldsymbol{\Phi} = 
                \begin{bmatrix}
                    \phi_1 & 0 \\
                    0 & \phi_1 \\
                    \phi_2 & 0 \\
                    0 & \phi_2 \\
                    \phi_3 & 0 \\
                    0 & \phi_3 \\
                    \phi_4 & 0 \\
                    0 & \phi_4 \\
                \end{bmatrix}
                \begin{bmatrix}
                    \phi_1 & 0 & \phi_2 & 0 & \phi_3 & 0 & \phi_4 & 0 \\
                    0 & \phi_1 & 0 & \phi_2 & 0 & \phi_3 & 0 & \phi_4 \\
                \end{bmatrix}
            \end{equation}
            and $\rho$ is the area of the element.
            As a result, we can use the expressions for the shape functions as well as $det(\textbf{J})$ from Eq. (\ref{Eq: detJ}) to calculate all sixty-four components of the mass matrix for the given element.
            It turns out, however, that only ten of these components are unique, so we can simplify the mass matrix significantly as follows:
            \begin{equation}
                \textbf{M} = 
                \begin{bmatrix}
                    M_{11} & M_{13} & M_{15} & M_{17} \\
                    M_{13} & M_{33} & M_{35} & M_{37} \\
                    M_{15} & M_{35} & M_{55} & M_{57} \\
                    M_{17} & M_{37} & M_{57} & M_{77}
                \end{bmatrix}
            \end{equation}
            where
            \begin{equation}
                M_{ij} = \rho \int_{-1}^{1} \int_{-1}^{1} \left[\phi_i \phi_j \cdot det (\textbf{J}) \right] d\xi d\eta.
            \end{equation}
            
            After obtaining the cumulative force on each node in the element through Gaussian integration (see \ref{App: Gaussian integration}), we can then calculate the respective accelerations as follows:
            \begin{equation}
                \begin{bmatrix}
                    \ddot{\textbf{u}}_1 \\
                    \ddot{\textbf{u}}_2 \\
                    \ddot{\textbf{u}}_3 \\
                    \ddot{\textbf{u}}_4
                \end{bmatrix} = 
                \begin{bmatrix}
                    M_{11} & M_{13} & M_{15} & M_{17} \\
                    M_{13} & M_{33} & M_{35} & M_{37} \\
                    M_{15} & M_{35} & M_{55} & M_{57} \\
                    M_{17} & M_{37} & M_{57} & M_{77}
                \end{bmatrix}^{-1}
                \begin{bmatrix}
                    \textbf{f}_1 \\
                    \textbf{f}_2 \\
                    \textbf{f}_3 \\
                    \textbf{f}_4
                \end{bmatrix}.
            \end{equation}
            In this work, we use the lumped mass matrix approximation, and specifically, the row-sum method. 
            Hence, we can further simplify our calculations and sum the rows of the mass matrix such that 
            \begin{align}
                M_1 &= M_{11} + M_{13} + M_{15} + M_{17} \\
                M_2 &= M_{13} + M_{33} + M_{35} + M_{37} \\
                M_3 &= M_{15} + M_{35} + M_{55} + M_{57} \\
                M_4 &= M_{17} + M_{37} + M_{57} + M_{77}.
            \end{align}
            Therefore, we arrive at the final result for the accelerations of the four nodes:
            \begin{equation}
                \begin{bmatrix}
                    \ddot{\textbf{u}}_1 \\
                    \ddot{\textbf{u}}_2 \\
                    \ddot{\textbf{u}}_3 \\
                    \ddot{\textbf{u}}_4
                \end{bmatrix} = 
                \begin{bmatrix}
                    \textbf{f}_1 / M_1 \\
                    \textbf{f}_2 / M_2 \\
                    \textbf{f}_3 / M_3 \\
                    \textbf{f}_4 / M_4
                \end{bmatrix}.
            \end{equation}
            
            For the sake of completeness, we provide the expressions for the ten unique components of the two-dimensional mass matrix below:
            \begin{align}
                M_{11} &= \frac{\rho}{36} \left[(3x_1 - x_3)(y_2 - y_4) + x_2(2y_4 + y_3 - 3y_1) + x_4(3y_1 - 2y_2 - y_3) \right] \\
                M_{13} &= \frac{\rho}{72} \left[x_1(3y_2 - y_3 - 2y_4) + x_2(y_4 + 2y_3 - 3y_1) + x_3(y_1 - 2y_2 + y_4) + x_4(2y_1 - y_2 - y_3) \right] \\
                M_{15} &= \frac{\rho}{72} \left[(x_1 - x_3)(y_2 - y_4) - (x_2 - x_4)(y_1 - y_3) \right] \\
                M_{17} &= \frac{\rho}{72} \left[x_1(2y_2 + y_3 - 3y_4) + x_2(y_4 + y_3 - 2y_1) - x_3(y_1 + y_2 - 2y_4) + x_4(3y_1 - y_2 - 2y_3) \right] \\
                M_{33} &= \frac{\rho}{36} \left[x_1(3y_2 - 2y_3 - y_4) + 3x_2(y_3 - y_1) + x_3(2y_1 - 3y_2 + y_4) + x_4(y_1 - y_3) \right] \\
                M_{35} &= \frac{\rho}{72} \left[x_1(2y_2 - y_3 - y_4) - x_2(2y_1 - 3y_3 + y_4) + x_3(y_1 - 3y_2 + 2y_4) + x_4(y_1 + y_2 - 2y_3) \right] \\
                M_{37} &= \frac{\rho}{72} \left[(x_1 - x_3)(y_2 - y_4) - (x_2 - x_4)(y_1 - y_3) \right] \\
                M_{55} &= \frac{\rho}{36} \left[(x_1 - 3x_3)(y_2 - y_4) - x_2(y_1 - 3y_3 + 2y_4) + x_4(y_1 + 2y_2 - 3y_3) \right] \\
                M_{57} &= \frac{\rho}{72} \left[x_1(y_2 + y_3 - 2y_4) - x_2(y_1 - 2y_3 + y_4) - x_3(y_1 + 2y_2 - 3y_4) + x_4(2y_1 + y_2 - 3y_3) \right] \\
                M_{77} &= \frac{\rho}{36} \left[x_1(y_2 + 2y_3 - 3y_4) + x_2(y_3 - y_1) - x_3(2y_1 + y_2 - 3y_4) + 3x_4(y_1 - y_3) \right]
            \end{align}
            As can be seen, each of these terms is strictly a function of the four nodal positions of the element in the global coordinate system as well as the density $\rho$. 
            Thus, assuming that we know the global coordinates, we can calculate each component of the mass matrix and thereby obtain the acceleration of each node.
            
        \subsection{Gaussian integration} \label{App: Gaussian integration}
\begin{figure}[htpb]
                \centering
                \includegraphics[width=0.55\textwidth]{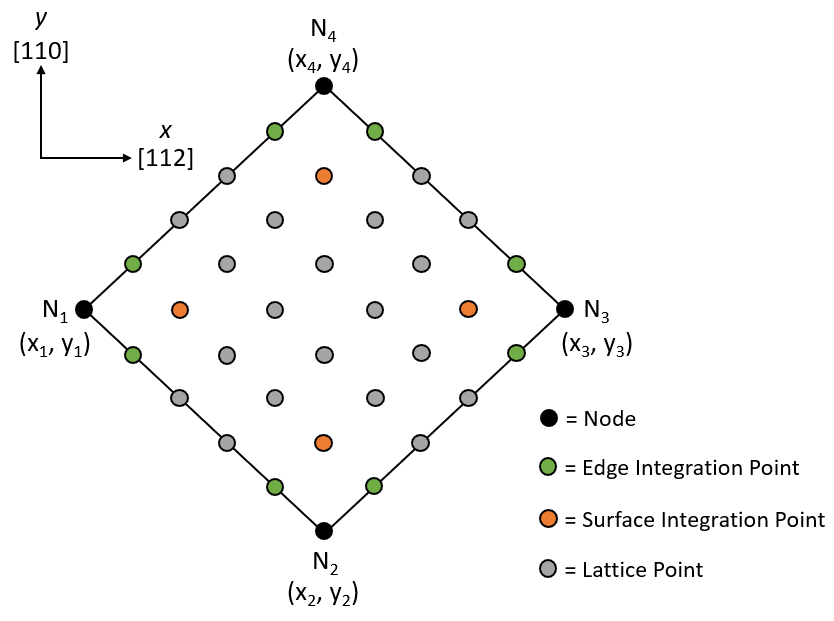}
                \caption{Two-dimensional CAC coarse-scaled element. 
                Nodes are shown in black, edge integration points are shown in green, surface integration points are shown in orange, and lattice points are shown in grey.}
                \label{Fig:2DCACElement}
            \end{figure}

            For this work, we calculate the internal force density using Gaussian integration, so we now elaborate on this method for a 2D coarse-scaled element.
            In Gaussian integration, the elemental forces are approximated by the forces at both the nodes as well as the integration points.
            Thus, while more complex to implement, Gaussian integration typically results in more accurate force calculations when using complex geometries or large elements.
            For all of our simulations, we use twelve-point Gaussian integration such that each element, in addition to the four nodes, contains twelve integration points.
            These integration points are chosen such that there are two along each edge of the element and four on the interior surface as seen in Fig. \ref{Fig:2DCACElement}.
            In particular, both the edge and surface integration points are chosen to be equal to the lattice points which directly neighbor the nodes, and this is comparable to techniques used in other multiscale schemes such as cluster-QC \cite{Knap2001Analysis}.
            
            The forces within the 2D element are thus split into three distinct parts associated with the (1) nodes, (2) edges, and (3) surfaces as seen in the equation below:
            \begin{equation} \label{Eq: GaussianTotalForce}
                \textbf{F}_{int} = \textbf{FN}_{int} + \textbf{FE}_{int} + \textbf{FS}_{int}.
            \end{equation}
            The first term in Eq. (\ref{Eq: GaussianTotalForce}) represents the forces at the nodes and is calculated as follows:
            \begin{align} \label{Eq: NodalTotalForce}
                \textbf{FN}_{int} &= w_{N_1} \boldsymbol{\Phi}(N_1) \textbf{f}_{int}(N_1) + w_{N_2} \boldsymbol{\Phi}(N_2)\textbf{f}_{int}(N_2) + w_{N_3} \boldsymbol{\Phi}(N_3) \textbf{f}_{int}(N_3) + w_{N_4} \boldsymbol{\Phi}(N_4)\textbf{f}_{int}(N_4) \nonumber \\[1ex]
                &= \begin{bmatrix}
                    \phi_1(N_1) \\[0.5ex]
                    \phi_2(N_1) \\[0.5ex]
                    \phi_3(N_1) \\[0.5ex]
                    \phi_4(N_1)
                \end{bmatrix}
                \textbf{f}_{int}(N_1) +
                \begin{bmatrix}
                    \phi_1(N_2) \\[0.5ex]
                    \phi_2(N_2) \\[0.5ex]
                    \phi_3(N_2) \\[0.5ex]
                    \phi_4(N_2)
                \end{bmatrix}
                \textbf{f}_{int}(N_2) + 
                \begin{bmatrix}
                    \phi_1(N_3) \\[0.5ex]
                    \phi_2(N_3) \\[0.5ex]
                    \phi_3(N_3) \\[0.5ex]
                    \phi_4(N_3)
                \end{bmatrix}
                \textbf{f}_{int}(N_3) +
                \begin{bmatrix}
                    \phi_1(N_4) \\[0.5ex]
                    \phi_2(N_4) \\[0.5ex]
                    \phi_3(N_4) \\[0.5ex]
                    \phi_4(N_4)
                \end{bmatrix}
                \textbf{f}_{int}(N_4) \nonumber \\[1ex]
                &= \begin{bmatrix}
                    1 \\[0.5ex]
                    0 \\[0.5ex]
                    0 \\[0.5ex]
                    0
                \end{bmatrix}
                \textbf{f}_{int}(x_1, y_1) +
                \begin{bmatrix}
                    0 \\[0.5ex]
                    1 \\[0.5ex]
                    0 \\[0.5ex]
                    0
                \end{bmatrix}
                \textbf{f}_{int}(x_2, y_2) + 
                \begin{bmatrix}
                    0 \\[0.5ex]
                    0 \\[0.5ex]
                    1 \\[0.5ex]
                    0
                \end{bmatrix}
                \textbf{f}_{int}(x_3, y_3) +
                \begin{bmatrix}
                    0 \\[0.5ex]
                    0 \\[0.5ex]
                    0 \\[0.5ex]
                    1
                \end{bmatrix}
                \textbf{f}_{int}(x_4, y_4) \nonumber \\[1ex]
                &= \begin{bmatrix}
                    \textbf{f}_{int}(x_1, y_1) \\[0.5ex]
                    \textbf{f}_{int}(x_2, y_2) \\[0.5ex]
                    \textbf{f}_{int}(x_3, y_3) \\[0.5ex]
                    \textbf{f}_{int}(x_4, y_4)
                \end{bmatrix}
            \end{align}
            where we note that all of the weights equal one.
            Additionally, each shape function equals one at its nodal location and zero everywhere else. 
            Equation (\ref{Eq: NodalTotalForce}) would be the only force used in nodal integration -- a technique which effectively does not alter the forces obtained using the interatomic potential function and relative displacement of particles.
            Although nodal integration is more computationally efficient, it is less robust than Gaussian integration and only accurate for simple geometries and relatively small elements, so it is not used in this work.

            The second two forces in Eq. (\ref{Eq: GaussianTotalForce}) are given as follows:
            \begin{align}
                \textbf{FE}_{int} &= \sum_{j = 1}^{8} (w_{j,x} \cdot w_{j,y}) \boldsymbol{\Phi}(j) \textbf{f}_{int}(j) \label{Eq: ElementEdgeForces} \\
                \textbf{FS}_{int} &= \sum_{j = 9}^{12} (w_{j,x} \cdot w_{j,y}) \boldsymbol{\Phi}(j) \textbf{f}_{int}(j) \label{Eq: ElementSurfaceForces}.
            \end{align}
            In Eqs. (\ref{Eq: ElementEdgeForces}) and (\ref{Eq: ElementSurfaceForces}), the summations occur over the eight edge integration points and four surface integration points respectively.
            Furthermore, the terms $w_{j,x}$ and $w_{j,y}$ are the weights of the integration points along the $x$ and $y$ directions.
            Finally, $\boldsymbol{\Phi}(j)$ is the shape function vector at the given integration point while $\textbf{f}_{int}(j)$ is the force of the integration point obtained through the potential function.
            For the sake simplicity, we do not write out the full expressions of these terms, but the expansion would be similar to that shown for the nodal forces in Eq. (\ref{Eq: NodalTotalForce}).
           
\end{document}